\documentclass[a4paper,fleqn,usenatbib,useAMS]{mnras}

\usepackage{graphicx}  
\usepackage{amsmath}  
\usepackage{amssymb}  
\usepackage{multicol}        
\usepackage{bm}    
\usepackage{pdflscape}  
\usepackage{multirow}
\usepackage{xcolor}
\usepackage{hyperref}
\usepackage{booktabs}
\usepackage{subfigure}
\usepackage{xspace}
\usepackage{tikz}
\usepackage[para,online,flushleft]{threeparttable}
\usepackage[colorinlistoftodos,prependcaption,textsize=tiny]{todonotes}
\usepackage{xspace}
\usepackage{orcidlink}

\newcommand{\fex}{\xspace f_{\rm ex-situ}} 
\newcommand{\mq}{{\rm Q}} 
\newcommand{\mstar}{\xspace M_{\rm star}} 
\newcommand{\mbh}{\xspace M_{\rm BH}} 
\newcommand{\kpc}{\xspace \text{kpc}} 
\newcommand{\msun}{\mathrm M_\odot} 
\newcommand{\orcid}[1]{\href{https://orcid.org/#1}{\textcolor[HTML]{A6CE39}{\aiOrcid}}}

\def\specialname[#1]{\textbf{\textsc{#1}}}

\def\apj{ApJ}
\def\aap{A\& A}
\def\apjl{ApJL}
\def\apjs{ApJS}
\def\mnras{MNRAS}

\def\aj{AJ}
\def\araa{ARA\& A}

\definecolor{lime}{HTML}{A6CE39}

\usepackage[T1]{fontenc}
\usepackage{ae,aecompl}
\usepackage{newtxtext,newtxmath}

\title[Forged in Quenching]{
  Forged in Quenching: Morphological Transformation across Star-forming and Quiescent Galaxies in EAGLE
}
\author[Wang et al.]{Kai Wang,$^{1,2}$\thanks{wkcosmology@gmail.com}\orcidlink{0000-0002-3775-0484}
  Carlton Baugh,$^{1,3}$\orcidlink{0000-0002-9935-9755}
  Sownak Bose,$^{1}$\orcidlink{0000-0002-0974-5266}
  Shaun Cole,$^{1}$\orcidlink{0000-0002-5954-7903}
  Carlos S. Frenk,$^{1}$\orcidlink{0000-0002-2338-716X}
  Hao Fu,$^{4}$\orcidlink{0009-0002-8051-1056}
  \newauthor
  Cedric Lacey,$^{1}$\orcidlink{0000-0001-9016-5332}
  Shengdong Lu,$^{1}$\orcidlink{0000-0002-6726-9499}
  Aaron Ludlow,$^{5}$\orcidlink{0000-0001-6119-4871}
  Peder Norberg,$^{1,2}$\orcidlink{0000-0002-5875-0440}
  Yingjie Peng,$^{6, 7}$\orcidlink{0000-0003-0939-9671}
  \newauthor
  Katy L. Proctor,$^{8}$\orcidlink{0009-0003-1836-7169}
  Isabel Santos-Santos,$^{9}$
  Francesco Shankar,$^{10}$\orcidlink{0000-0001-8973-5051}
  Tom Theuns,$^{1}$\orcidlink{0000-0002-3790-9520}
  Enci Wang,$^{11, 12}$\orcidlink{0000-0003-1588-9394}
  \newauthor
  Tao Wang,$^{13, 14}$\orcidlink{0000-0002-2504-2421}
  and Vivienne Wild$^{15}$\orcidlink{0000-0002-8956-7024}
  \\
  $^{1}$Institute for Computational Cosmology, Department of Physics, Durham University, South Road, Durham, DH1 3LE, UK\\
  $^{2}$Centre for Extragalactic Astronomy, Department of Physics, Durham University, South Road, Durham DH1 3LE, UK\\
  $^{3}$Institute for Data Science, Durham University, South Road, Durham DH1 3LE, UK\\
  $^{4}$Center for Astronomy and Astrophysics and Department of Physics, Fudan University, Shanghai 200438, China\\
  $^{5}$International Centre for Radio Astronomy Research (ICRAR), University of Western Australia, Crawley, WA 6009, Australia\\
  $^{6}$Department of Astronomy, School of Physics, Peking University, Beijing 100871, China\\
  $^{7}$Kavli Institute for Astronomy and Astrophysics, Peking University, Beijing 100871, China\\
  $^{8}$The Oskar Klein Centre, Department of Physics, Stockholm University, AlbaNova University Center, 106 91 Stockholm, Sweden\\
  $^{9}$Leibniz-Institut für Astrophysik Potsdam (AIP), An der Sternwarte 16, 14482 Potsdam, Germany\\
  $^{10}$School of Physics and Astronomy, University of Southampton, Highfield, Southampton SO17 1BJ, UK\\
  $^{11}$Department of Astronomy, University of Science and Technology of China, Hefei 230026, China\\
  $^{12}$School of Astronomy and Space Science, University of Science and Technology of China, Hefei 230026, China\\
  $^{13}$School of Astronomy and Space Science, Nanjing University, Nanjing, Jiangsu 210093, China\\
  $^{14}$Key Laboratory of Modern Astronomy and Astrophysics, Nanjing University, Ministry of Education, Nanjing 210093, China\\
  $^{15}$School of Physics \& Astronomy, University of St Andrews, North Haugh, St Andrews KY16 9SS, UK
}

\date{Last updated 2025 May 22; in original form 2025 May 5}

\pubyear{2025}

\begin{document}
\label{firstpage}
\pagerange{\pageref{firstpage}--\pageref{lastpage}}
\maketitle


\begin{abstract}
  The connection between morphology and quenching in central galaxies is well established, but its physical origin remains widely debated.
  We address this by tracing the main progenitor branches of $z=0$ star-forming and quiescent central galaxies in the EAGLE cosmological simulation from $z\gtrsim4$.
  Their disc-to-total ratio and triaxiality tracks are indistinguishable until $z\approx 1$--$2$, when both diverge concurrently with the onset of quenching, whereas the size and supermassive black hole (SMBH) mass differences are established earlier.
  We identify four physically distinct channels linking galaxy morphology and quenching.
  First, mergers cause size growth, rotation suppression, triaxiality increase, and SMBH growth, with the accumulated SMBH mass subsequently causes the quenching of galaxies.
  Second, with merger history controlled, galaxy morphology modulates SMBH growth throughout the star-forming phase: compact, dispersion-dominated galaxies grow their SMBHs faster and are preferentially quenched, producing the size and morphology differences between star-forming and quiescent galaxies.
  Third, at fixed stellar mass and SMBH mass, compactness further facilitates the quenching of galaxies.
  Fourth, disc instability transforms compact oblate discs into prolate systems, with substantial size growth and suppressed rotation but negligible stellar mass growth.
  This secular channel contributes about half of the prolate galaxy population around $M_{\rm star}\approx 10^{10.6}\,\rm M_\odot$.
  Prior to quenching, the progenitors of quiescent galaxies already have smaller sizes, lower disc-to-total ratios, and more massive SMBHs than star-forming galaxies at the same epoch, by amounts comparable to their differences at $z=0$.
  Morphology therefore plays an active role in growing the SMBH and quenching the galaxy, rather than being passively inherited through progenitor bias.
  \end{abstract}


\section{Introduction}
\label{sec:introduction}

The galaxy population in the local Universe is characterised by a striking bimodality.
Galaxies segregate into a blue cloud of actively star-forming systems and a red sequence of quiescent galaxies, with a relatively sparsely populated green valley in between \citep[e.g.][]{baldryGalaxyBimodalityStellar2006, pengMassEnvironmentDrivers2010, wangSDSSIVMaNGAStar2018}.
This bimodality manifests not only in star formation activity and colour, but also in morphology \citep[see][]{blantonPhysicalPropertiesEnvironments2009}: star-forming galaxies tend to be extended, rotationally supported discs, while quiescent galaxies tend to be compact, dispersion-supported spheroids \citep{vanderwel3DHSTCANDELSEvolution2014}.
This morphology--star formation correlation is remarkably tight across a wide range of redshifts and environments \citep[e.g.][]{conseliceEvolutionGalaxyStructure2014}, and poses a fundamental question: are the processes that quench star formation the same as those that reshape galaxies, or do morphological transformation and quenching merely correlate because they share a common origin?

The canonical answer invokes galaxy mergers as the common driver of both. Mergers destroy rotational support, produce compact pressure-supported remnants, and drive gas towards the centre, fuelling rapid supermassive black hole (SMBH) growth and the AGN feedback that subsequently suppresses star formation \citep[see][]{hernquistTidalTriggeringStarbursts1989, mooreGalaxyHarassmentEvolution1996, naabStatisticalPropertiesCollisionless2003, dimatteoEnergyInputQuasars2005, hopkinsUnifiedMergerdrivenModel2006, bowerBreakingHierarchyGalaxy2006, crotonManyLivesActive2006}.
In this picture, morphological transformation and quenching are two faces of the same merger-driven process, with the degree of structural transformation depending sensitively on the mass ratio, orbital properties, and the gas fraction of the merging systems \citep[see][]{hopkinsHowDisksSurvive2009, zengFormationMassiveDisc2021, proctorWeakConnectionStellar2025}.

Observations show that blue star-forming galaxies are more extended than their red quiescent counterparts at fixed stellar mass \citep{vanderwel3DHSTCANDELSEvolution2014, wangDearthDifferencesCentral2020, vanderwelStellarHalfmassRadii2024, chen2026masssize}.
A critical tension arises when this observed difference is interpreted through models in which mergers transform star-forming disc galaxies into quiescent spheroids.
Mergers drive size growth \citep{coleHierarchicalGalaxyFormation2000, naabMinorMergersSize2009, hopkinsDiscriminatingPhysicalProcesses2010, oserCosmologicalSizeVelocity2012}, and \citet{wang26} find that a single merger event increases galaxy size by $\approx0.03$--$0.1$ dex on average, with cumulative mergers driving larger size growth.
A quiescent population assembled through this channel would therefore be more extended than its star-forming counterparts at fixed stellar mass, the opposite of what is observed.
The tension points to an additional physical mechanism connecting galaxy structure and quenching that operates independently of the merger channel.
Such a mechanism is captured in an empirical framework in which galaxy size is a key secondary parameter governing SMBH growth \citep[e.g.][]{vandokkumFormingCompactMassive2015, krajnovicTwoChannelsSupermassive2018, chenQuenchingContestGalaxy2020}: at fixed stellar mass, more compact galaxies have higher central stellar densities and grow their SMBHs more efficiently, so they are more likely to be quenched by the resulting more powerful AGN feedback, while larger galaxies grow their SMBHs more slowly and are less likely to reach the SMBH mass needed for AGN feedback to overcome the gravitational binding energy of the halo \citep{whitakerPredictingQuiescenceDependence2017, chenQuenchingContestGalaxy2020}.
In this picture, the size difference between star-forming and quiescent galaxies is not a consequence of quenching but a structural \textit{precondition} that determines which galaxies quench first.

These ideas have been examined in cosmological simulations.
In IllustrisTNG \citep{pillepichSimulatingGalaxyFormation2018}, \citet{tacchellaMorphologyStarFormation2019} found that galaxy morphology, as measured by the spheroid-to-total ratio, is largely established during the star-forming phase and does not change significantly at the time of quenching.
They interpreted the morphological difference between star-forming and quiescent galaxies at $z = 0$ as a consequence of \textit{progenitor bias} \citep{carolloZURICHENVIRONMENTALSTUDY2013, lillySURFACEDENSITYEFFECTS2016}: quiescent galaxies preserve the morphology they had at the moment of quenching, which reflects the higher spheroid-to-total ratios characteristic of earlier cosmic epochs.
However, \citet{genelSizeEvolutionStarforming2018} showed that the size offset between star-forming and quiescent galaxies in IllustrisTNG is already in place prior to quenching and continues to grow thereafter, suggesting that progenitor bias alone cannot account for the structural differences between the two populations.
In EAGLE \citep{schayeEAGLEProjectSimulating2015, crainEAGLESimulationsGalaxy2015}, \citet{correaRelationGalaxyMorphology2017} showed that kinematic morphology is tightly correlated with galaxy colour across the full population, with the red sequence predominantly occupied by dispersion-dominated systems and the blue cloud by rotationally supported discs.
\citet{correaOriginRedsequenceGalaxy2019} further traced the progenitors of galaxies that are on the red-sequence at $z=0$ and found that the epoch at which galaxies join the red sequence depends on their morphological type at $z=0$, with elliptical centrals quenching earlier and their reddening correlated in time with peak AGN activity, whereas morphological and colour changes are not strongly correlated in time and its physical origin remains unidentified \citep[see also][]{daviesQuenchingMorphologicalEvolution2020}.
Together, these results establish that morphological transformation and quenching share common physical drivers, yet the nature of these drivers has not been clearly identified.

Moreover, these studies are subject to several important limitations.
First, they characterise morphology through spherically averaged structural parameters such as size, spheroid-to-total ratio, concentration index, or corotation parameter, which are insensitive to the full three-dimensional distribution of stellar particles.
\citet{thobRelationshipMorphologyKinematics2019} demonstrated in EAGLE that three-dimensional shape parameters correlate more strongly with galaxy colour than any single conventional structural parameter, yet the physical origin of this correlation remains unexplored.
Second, the role of mergers is not actively controlled, making it difficult to uncover the processes governing the morphology--quenching connection beyond merger activity.
Third, post-quenching morphological evolution is implicitly assumed to be negligible once galaxies are quenched and free of mergers, yet \citet{genelSizeEvolutionStarforming2018} found continued size growth in IllustrisTNG after quenching; whether analogous secular structural evolution operates in merger-poor galaxies, and through what mechanism, has not been tested \citep[see also][]{trayfordStarFormationRate2019}.

In this paper we revisit the relation between morphology and galaxy quenching using the EAGLE cosmological simulation \citep{schayeEAGLEProjectSimulating2015, crainEAGLESimulationsGalaxy2015, mcalpineEagleSimulationsGalaxy2016}.
We select $\approx 2{,}200$ central galaxies at $z=0$ with $M_\mathrm{star} \geqslant 10^{10}\,\msun$ and trace their evolution back in time along their main progenitor branch up to  $z \gtrsim 4$.
By using the ex-situ stellar mass fraction as a proxy for merger history and examining the relationship between galaxy structure and SMBH growth, we identify four physically distinct channels through which morphological transformation and quenching are connected.
The paper is organised as follows.
In \S~\ref{sec:simulation_data} we describe the simulation and morphological measurements.
In \S~\ref{sec:morphology_evolution} we trace the morphological evolution of star-forming and quiescent galaxies along their main progenitor branches and characterise the divergence between the two populations.
In \S~\ref{sec:associate_quenching_morphology} we identify the physical drivers of this divergence, decomposing the morphology--quenching connection into distinct channels through merger activity, morphology-modulated SMBH growth, size-dependent AGN feedback quenching, and post-quenching secular evolution.
In \S~\ref{sec:prolate_galaxies_formed_out_of_disc_instability} we characterise the demographics of the prolate galaxy population and explore the physical origin of prolate morphologies.
We discuss the implications of our findings in \S~\ref{sec:discussion} and summarise our conclusions in \S~\ref{sec:summary}.

\section{Simulation data}
\label{sec:simulation_data}

This work makes use of the EAGLE suite of cosmological hydrodynamical simulations \citep{schayeEAGLEProjectSimulating2015, crainEAGLESimulationsGalaxy2015, mcalpineEagleSimulationsGalaxy2016}.
Specifically, we use the flagship run, \textsc{Ref-L100N1504}, which follows the evolution of baryons and dark matter within a periodic cubic volume of side length of 100 cMpc.
The simulation employs $1504^3$ dark matter particles and an equal number of initial gas particles, with dark matter and initial gas particle masses of $m_{\rm DM}=9.70\times 10^6\,\rm M_\odot$, $m_{\rm gas} = 1.81 \times 10^6\,\mathrm{M_\odot}$, respectively.
The Plummer-equivalent gravitational softening length is kept fixed in comoving units at $2.66\,\mathrm{ckpc}$ down to $z = 2.8$, and thereafter fixed in proper units at $\epsilon = 0.70\,\mathrm{pkpc}$.

Haloes are identified with a friends-of-friends algorithm and gravitationally bound substructures are found with \textsc{SUBFIND} \citep{springelPopulatingClusterGalaxies2001}.
The central galaxy of each FoF halo is defined as the subhalo containing the particle with the minimum gravitational potential, which also defines the halo centre.
Stellar masses, $M_{\rm star}$, and other galaxy properties are measured within a 3D aperture of radius 30 pkpc centred on each subhalo.
Merger trees are constructed with the \textsc{D-Trees} algorithm \citep{jiangNbodyDarkMatter2014}, and the main progenitor branch of each subhalo is defined following the EAGLE convention as the branch with the highest integrated total mass \citep{quChronicleGalaxyMass2017}.

We select all central galaxies at $z=0$ with stellar mass $M_{\rm star}\geq10^{10}\,\rm M_\odot$, well above the threshold below which spurious numerical heating significantly affects the structural and kinematic properties of galaxies in EAGLE \citep{ludlowSpuriousHeatingStellar2023}, yielding a sample of $\sim2200$ galaxies.
Restricting to central galaxies minimises the influence of environmental processes; we further exclude galaxies that have spent more than one snapshot as a satellite to remove potential backsplash galaxies \citep{wangDissectTwohaloGalactic2023}.
Each galaxy is traced back along its main progenitor branch to a minimum progenitor stellar mass of $10^9\,\rm M_\odot$, corresponding to at least $\sim 500$ stellar particles.
Galaxies with ${\rm sSFR}\equiv {\rm SFR}/M_{\rm star} < t_{\rm age}(0)/t_{\rm age}(z) \times 10^{-11}{\rm yr}^{-1}$ are deemed to be quiescent, and those above this threshold as star-forming.
Here $t_{\rm age}(z)$ is the age of the universe at redshift $z$.
For a given sample, the quiescent fraction $f_\mq$ is the number of quiescent galaxies divided by the total number of galaxies in that sample.

We adopt the morphological and kinematic properties presented in \citet{thobRelationshipMorphologyKinematics2019}.
All quantities are measured within a 30 pkpc spherical aperture.
The galaxy size, $r_{\rm star}$, is defined as the 3D half-mass radius of the bound stellar particles within that aperture.
We define the rotation axis as the direction of the stellar angular momentum within this aperture, and it is used to define the $z$-component of each particle's angular momentum $L_{z, i}$.
The disc-to-total stellar mass ratio, $\rm D/T$, adopts the standard estimator in which the bulge mass equals twice\footnote{This factor of two accounts for the assumption that the bulge component is symmetrically distributed about $L_{z,i}=0$.} the mass in counter-rotating stellar particles, i.e. $L_{z, i} < 0$,
\begin{equation}
  {\rm D/T} = 1 - \frac{2}{\sum_{i}m_i}\sum_{i, L_{z, i} < 0}m_i
\end{equation}
where $m_i$ is the mass for the $i$-th stellar particle.

Galaxy shape is quantified by fitting an ellipsoid to the stellar mass distribution using an iterative reduced inertia tensor.
The iteration begins from the set of stellar particles within a spherical aperture of 30 pkpc and stops when both $b/a$ and $c/a$ change by less than 1 per cent between iterations \citep[see][for details]{thobRelationshipMorphologyKinematics2019}.
The axis lengths $a\ge b\ge c$ are obtained from the eigenvalues of the converged tensor and define the flattening and triaxiality,
\begin{equation}
  \epsilon \equiv 1 - \frac{c}{a}, \qquad T\equiv \frac{a^2 - b^2}{a^2 - c^2}
\end{equation}
Here $\epsilon=0$ corresponds to a sphere, and low (high) values of $T$ correspond to oblate (prolate) ellipsoids.
\citet{forouharmorenoMorphologiesPresentdayGalaxies2026} show that these shape parameters converge for galaxies sampled by more than 100 stellar particles.
Our sample requires at least 500 stellar particles per galaxy and so lies comfortably above this limit.

Finally, we quantify the contribution of mergers to each galaxy's stellar mass assembly through the \textit{ex-situ} stellar mass fraction, $f_{\rm ex-situ}$, following the methodology of \citet{proctorWeakConnectionStellar2025}.
Each stellar particle present at $z=0$ is traced back to the first snapshot after its formation: if it is not bound to the main progenitor of the $z=0$ subhalo at that snapshot, i.e. it formed in a satellite of the main progenitor or in an external galaxy, the stellar particle is flagged as \textit{ex situ}.
The \textit{ex-situ} fraction is then the mass fraction of all such particles, and serves as a proxy for the integrated merger activity experienced by a galaxy over its lifetime.

\section{Morphology divergence between star-forming and quiescent galaxies}
\label{sec:morphology_evolution} 

\begin{figure*}
  \begin{center}
    \includegraphics[width=0.95\linewidth]{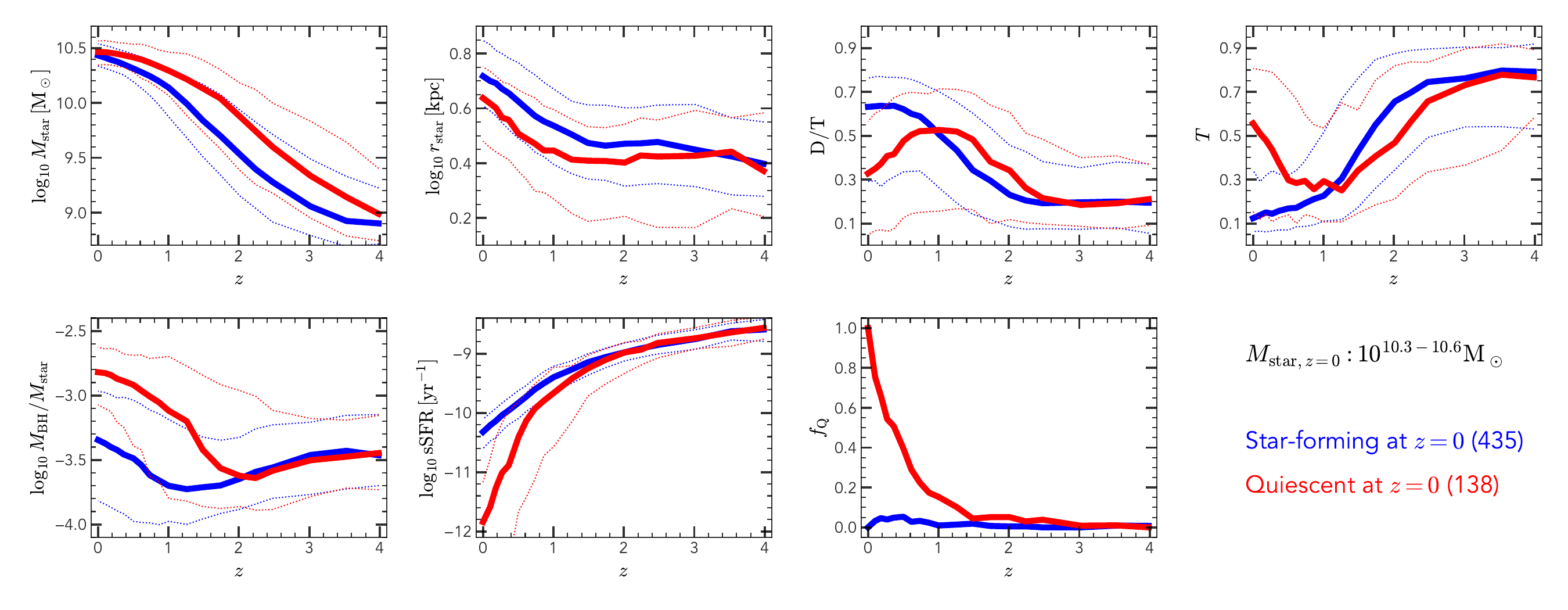}
  \end{center}
  \caption{
    Evolution of stellar mass ($M_{\rm star}$), size ($r_{\rm star}$), disc-to-total ratio ($\rm D/T$), triaxiality ($T$), SMBH-to-stellar mass ratio ($M_{\rm BH}/M_{\rm star}$), sSFR, and quiescent fraction ($f_{\rm Q}$) as a function of redshift along the main progenitor branch, for star-forming and quiescent central galaxies selected at $z=0$ with $M_{{\rm star},z=0}=10^{10.3}-10^{10.6}\,{\rm M}_\odot$.
    Thick solid lines show the median evolution and thin dotted lines show the $16^{\rm th}$--$84^{\rm th}$ percentile range.
    Numbers in parentheses give the size of each subsample.
    The narrow $z=0$ mass selection combined with suppressed late-time stellar mass growth means quiescent progenitors are systematically more massive than star-forming progenitors at fixed redshift; the correspondingly higher $\rm D/T$ and lower $T$ of quiescent progenitors before $z\approx1.5$ reflect this mass offset rather than an intrinsic morphological difference.
    SMBH growth in quiescent progenitors becomes elevated shortly before $z\approx1.5$, followed by the onset of quenching and a concurrent decline in $\rm D/T$ and rise in $T$.
    Quiescent progenitors remain more compact than star-forming progenitors at fixed redshift since $z\approx3$, even while both populations are star-forming and the quiescent progenitors are more massive, indicating that the size difference is not a mass effect.
    These trends hold only at the population level; the progenitor distributions of star-forming and quiescent galaxies overlap substantially at all redshifts.
  }
  \label{fig:evolution_afo_mstar1}
\end{figure*}

\begin{figure*}
  \begin{center}
    \includegraphics[width=0.95\linewidth]{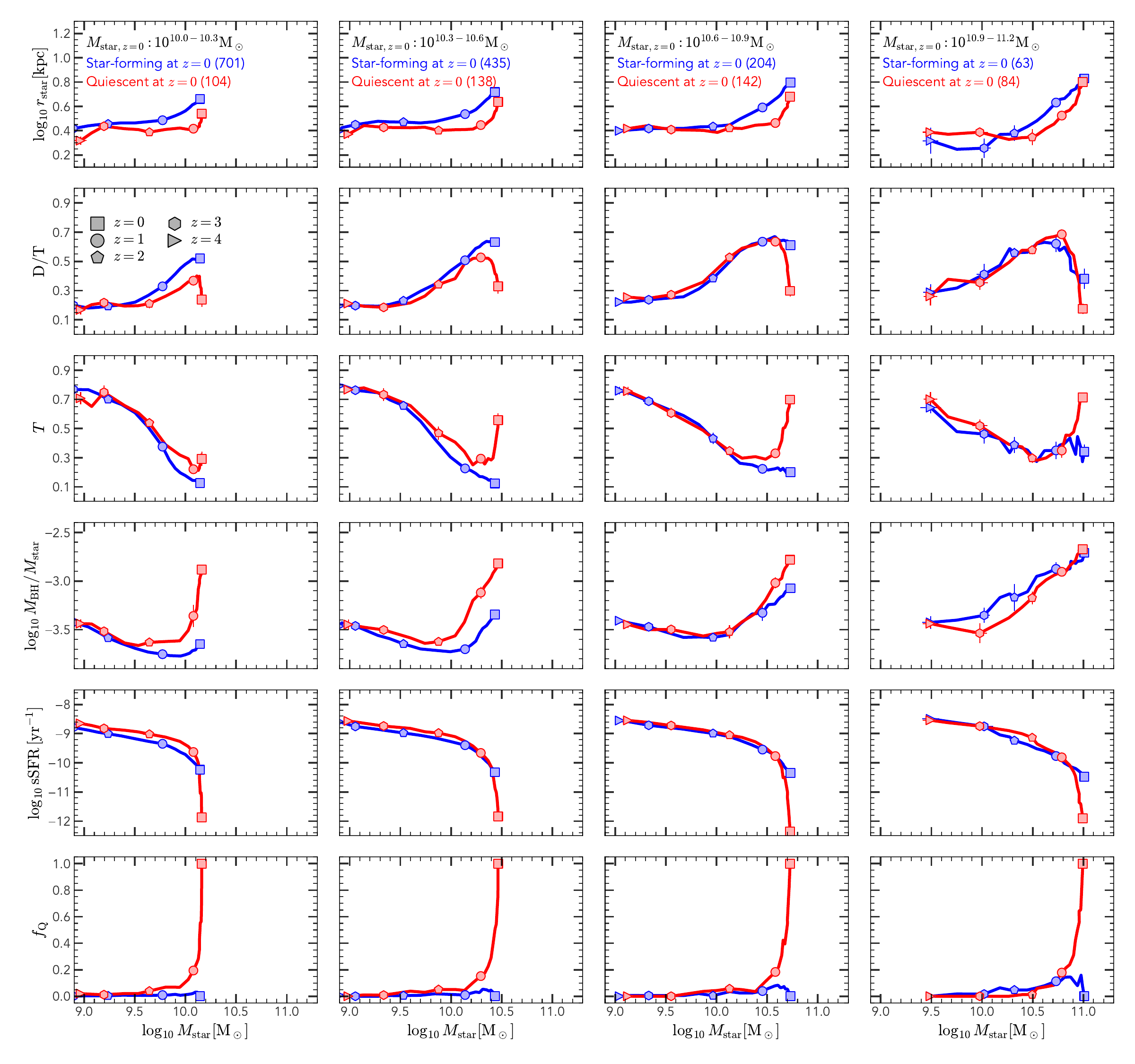}
  \end{center}
  \caption{
    Median properties of central galaxies along their main progenitor branches, shown as a function of the median progenitor stellar mass, $M_{\rm star}$, for four bins of $z=0$ stellar mass (columns, labelled at the top).
    From top to bottom, the panels show the galaxy size ($r_{\rm star}$), disc-to-total stellar mass ratio ($\rm D/T$), triaxiality ($T$), BH-to-stellar mass ratio ($\mbh/\mstar$), specific SFR,  and quiescent fraction ($f_\mq$).
    Blue and red curves correspond to galaxies classified as star-forming (${\rm sSFR} > 10^{-11}\,\rm yr^{-1}$) and quiescent (${\rm sSFR} < 10^{-11}\,\rm yr^{-1}$) at $z=0$, respectively.
    Different markers indicate different redshifts along each progenitor track, from $z=0$ to $z=4$, as indicated in the legend.
    Error bars show the uncertainty on the median estimated from 100 bootstrap samples.
    Numbers in parentheses give the size of each subsample.
    At fixed present-day stellar mass, quiescent galaxies exhibit systematically smaller sizes, lower $\rm D/T$, and markedly higher triaxiality than their star-forming counterparts, with the divergence between the two populations emerging at intermediate redshift
  }
  \label{fig:evolution_afo_mstar2}
\end{figure*}

Fig.~\ref{fig:evolution_afo_mstar1} shows the evolution of stellar mass ($M_{\rm star}$), galaxy size ($r_{\rm star}$), disc-to-total ratio ($\rm D/T$), triaxiality ($T$), SMBH-to-stellar mass ratio ($M_{\rm BH}/M_{\rm star}$), specific star formation rate (sSFR), and quiescent fraction ($f_\mq$) along the main progenitor branch for star-forming and quiescent central galaxies selected at $z=0$ with $M_{{\rm star},z=0}=10^{10.3}-10^{10.6}\,{\rm M}_\odot$.
Because the two populations are selected within a narrow $z=0$ stellar mass range and quiescent galaxies undergo suppressed stellar mass growth at late times, quiescent progenitors are systematically more massive than star-forming progenitors at fixed redshift, and remain so out to $z\approx4$ \citep[see also][]{clauwensThreePhasesGalaxy2018, wangRelatingGalaxiesDifferent2023, wangTestingGalaxyFormation2025}.
Before $z\approx1.5$, quiescent progenitors accordingly have higher D/T and lower $T$ than their star-forming counterparts, consistent with the general trend of increasing D/T and declining $T$ with increasing progenitor mass established in Fig.~\ref{fig:evolution_afo_mstar1}.
SMBH growth in quiescent progenitors becomes elevated shortly before $z\approx1.5$, after which the population quenches rapidly; the rise in $f_\mq$ is accompanied by a decline in D/T and a rise in $T$, reversing the sense of the morphological offset between the two populations established at higher redshift.
Quiescent progenitors are more compact than star-forming progenitors at fixed redshift since $z\approx3$, despite being more massive throughout this period, indicating that the size difference between the two populations is not attributable to their mass difference.

The comparison in Fig.~\ref{fig:evolution_afo_mstar1}, however, cannot separate the effect of stellar mass from the effect of star formation status: because quiescent and star-forming progenitors are not compared at fixed stellar mass, the D/T and $T$ offsets before $z\approx1.5$ could equally be driven by the mass difference between the two populations rather than by any structural distinction tied to their eventual star formation status.
We isolate these two effects in Fig.~\ref{fig:evolution_afo_mstar2}, which shows the same set of quantities along the main progenitor branch, now split into five bins of present-day stellar mass so that star-forming and quiescent trajectories can be compared at fixed progenitor mass.
Both populations share similar evolutionary trajectories at $z\gtrsim1-2$ within each mass bin, so we first use this common early evolution to characterise the general pattern of morphological growth before examining where the two populations diverge.

At the lowest progenitor masses probed ($M_{\rm star} \approx 10^9\msun$), galaxies are compact ($r_{\rm star}\approx 2.5\,\kpc$), with low disc-to-total ratio (${\rm D/T}\approx 0.2$) and high triaxiality ($T\approx 0.7$), indicative of small, prolate, dispersion-dominated systems.
This is consistent with recent observational constraints from JWST.
\citet{pandyaGalaxiesGoingBananas2024} reconstructed the three-dimensional shapes of high-redshift galaxies and found effective radii in the range $2-3\,\kpc$ for galaxies with $10^9 \lesssim M_{\rm star}/\msun \lesssim 10^{10.5}\,\msun$, with a weak dependence on both stellar mass and redshift across the range $2\leq z \leq 8$ \citep[see also][]{zhangEvolutionGalaxyShapes2019}.
Their inferred 3D shape statistics further show that galaxies at $M_{\rm star} \approx 10^9\,\msun$ are predominantly prolate with a mean triaxiality of $T \approx 0.9$, declining to $T \approx 0.4$ at $M_{\rm star} \gtrsim 10^{10}\,\msun$, a mass--triaxiality relation that is itself only weakly redshift-dependent across the same redshift range.
The general behaviour seen in EAGLE at low progenitor masses is therefore in good agreement with these observational findings, lending confidence that the structural properties of low-mass galaxies in the simulation are physically realistic.

As galaxies grow in stellar mass, $r_{\rm star}$ remains approximately constant at $\approx 2.5\,\kpc$ while the morphology evolves substantially: around $M_{\rm star} \approx 10^{10}-10^{10.5}\msun$, disc-to-total ratio ($\rm D/T$) rises to $\approx 0.4-0.7$ and triaxiality ($T$) falls below $\approx 0.1-0.3$, marking the establishment of rotationally supported discs.
This transition is consistent across all present-day mass bins, suggesting it reflects a characteristic mass scale above which gas accretion is sufficiently smooth and sustained to build coherent angular momentum \citep[see also][]{tomassettiEvolutionGalaxyShapes2016, clauwensThreePhasesGalaxy2018, hopkinsWhatCausesFormation2023}.

The two populations follow indistinguishable trajectories in $\rm D/T$ and $T$ at fixed progenitor mass until $z \lesssim 1$, after which galaxies destined to become quiescent develop systematically lower $\rm D/T$ and markedly higher triaxiality, concurrent with the rapid rise in quiescent fraction shown in the bottom panel of Fig.~\ref{fig:evolution_afo_mstar2}.
The size divergence begins somewhat earlier, at $z \sim 2$, suggesting that galaxy size evolution is not purely mass-driven but also sensitive to redshift \citep[see][]{genelSizeEvolutionStarforming2018, roperFirstLightReionization2023}.
By $z=0$, quiescent galaxies reach median triaxiality of $T\approx 0.3$--$0.7$ across all mass bins, while star-forming galaxies maintain $T\approx 0.1$--$0.3$, consistent with predominantly oblate, disc-dominated configurations \citep{vanderwel3DHSTCANDELSEvolution2014}.
The progenitors of quiescent galaxies also develop elevated $\mbh/\mstar$ ratios at a similar epoch to the morphological divergence.
The sSFR and $f_\mq$ panels of Fig.~\ref{fig:evolution_afo_mstar2} show that the suppression of star formation and the rise of the quiescent fraction occur at the same progenitor mass scale as the morphological divergence, across all present-day mass bins.
This concurrency is broadly consistent with results from both IllustrisTNG \citep[see][]{genelSizeEvolutionStarforming2018, tacchellaMorphologyStarFormation2019} and previous work with EAGLE \citep[see][]{furlongSizeEvolutionNormal2017, daviesQuenchingMorphologicalEvolution2020}.
Together, these trends indicate that SMBH growth, morphological transformation, and quenching share common physical drivers.

\section{The relation between quenching and morphological transformation}
\label{sec:associate_quenching_morphology}

\subsection{Galaxy mergers}
\label{sub:merger_driver} 

\begin{figure*}
  \begin{center}
    \includegraphics[width=0.9\linewidth]{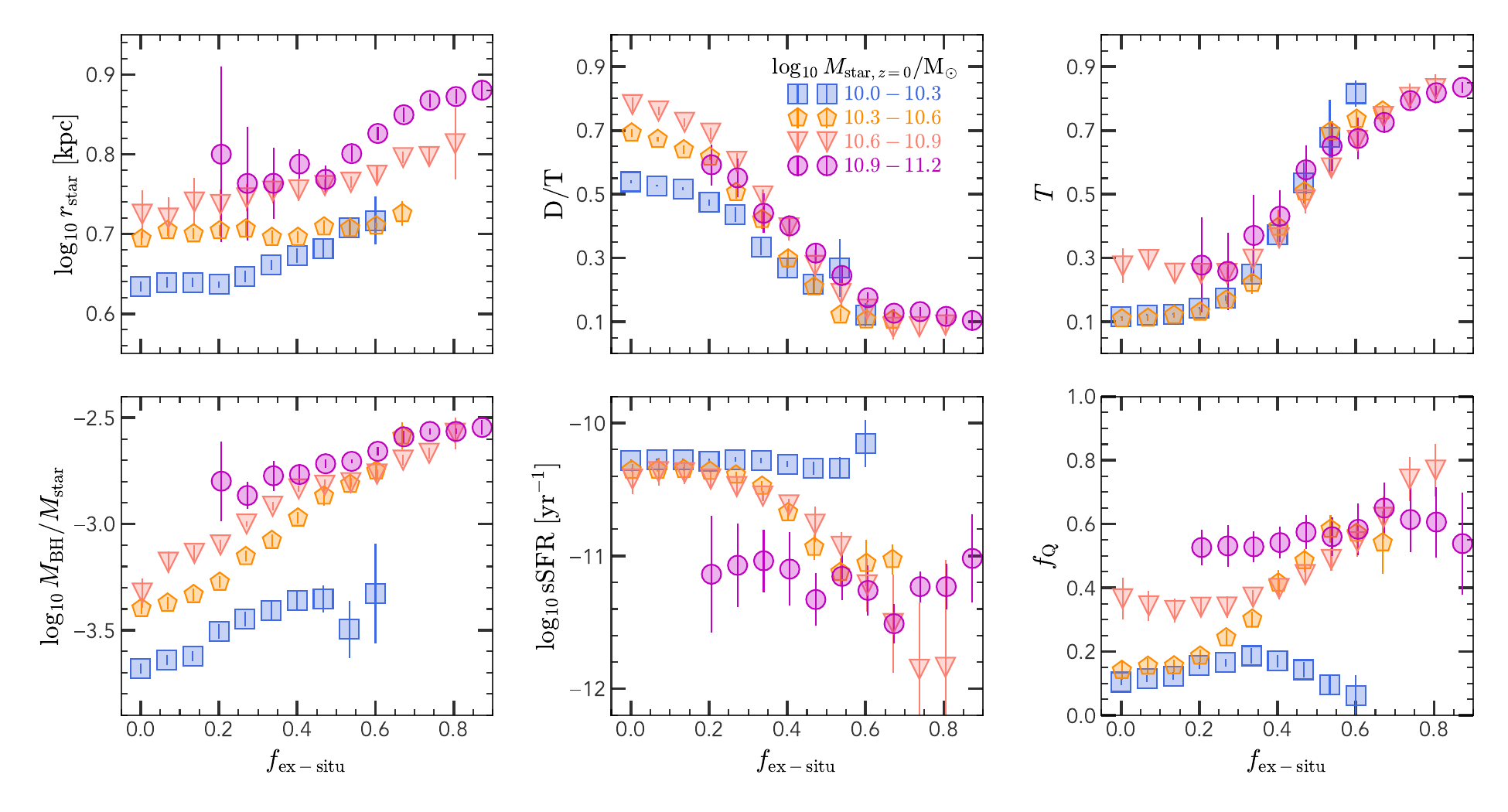}
  \end{center}
  \caption{
    Median galaxy properties at $z=0$ as a function of the ex-situ stellar mass fraction, $f_{\rm ex-situ}$, in four bins of present-day stellar mass (colours and markers as indicated in the legend).
    The six panels show: galaxy size ($r_{\rm star}$), disc-to-total ratio ($\rm D/T$), triaxiality ($T$), BH-to-stellar mass ratio ($M_{\rm BH}/M_{\rm star}$), sSFR, and quiescent fraction ($f_\mq$).
    Error bars show the uncertainty on the median estimated from 100 bootstrap samples.
    At fixed stellar mass, galaxies with higher $\fex$ are systematically larger, more triaxial, less disc-dominated, and, except in the lowest mass bin, more likely to be quiescent.
    Mergers therefore associate morphological transformation with quenching by driving both the structural change and the SMBH growth that ultimately quenches the galaxy.
  }
  \label{fig:morphology_merger}
\end{figure*}

\begin{figure}
  \begin{center}
    \includegraphics[width=0.9\linewidth]{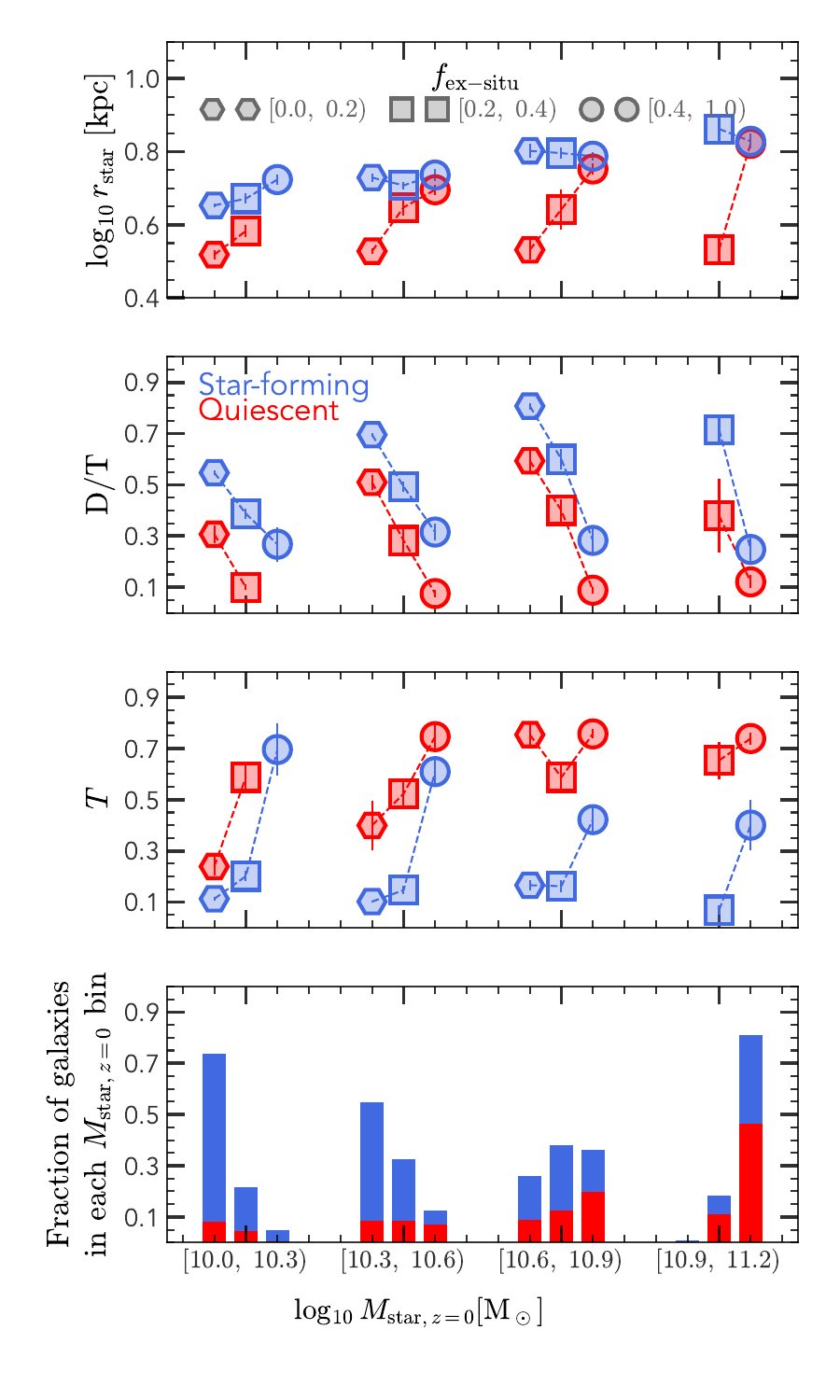}
  \end{center}
  \caption{
    Morphological properties of central galaxies at $z=0$ as a function of stellar mass, split simultaneously by star formation status (blue: star-forming; red: quiescent) and ex-situ stellar mass fraction.
    From top to bottom, the panels show galaxy size ($r_{\rm star}$), disc-to-total ratio ($\rm D/T$), triaxiality($T$), and the fractional distribution of galaxies.
    Error bars show the uncertainty on the median estimated from 100 bootstrap samples.
    Dashed lines connect the three $\fex$ bins at fixed stellar mass to guide the eye.
    The bottom panel shows the distribution of galaxies in three $\fex$ bins (from left to right) and in star-forming (blue) and quiescent (red) state, normalised in each $z=0$ stellar mass bin.
    At fixed stellar mass \textit{and} fixed $\fex$, quiescent galaxies remain systematically more compact, have lower $\rm D/T$, and higher triaxiality than their star-forming counterparts, especially for low-$\fex$ bin.
    This residual morphological difference, present even after controlling for merger history, demonstrates that \textit{mergers alone cannot account for the morphology-quenching connection}.
  }
  \label{fig:morphology_quenching_fexsitu}
\end{figure}

Mergers represent the canonical mechanisms of both morphological transformation and quenching in galaxy formation models, as they simultaneously destroy rotational support, trigger starbursts, fuel SMBH growth, and trigger AGN feedback \citep[see][]{hernquistTidalTriggeringStarbursts1989, coleHierarchicalGalaxyFormation2000, crotonManyLivesActive2006, bowerBreakingHierarchyGalaxy2006}.
Here we use the ex-situ stellar mass fraction, $\fex$, as a proxy for the integrated merger activity each galaxy has experienced over its lifetime \citep{davisonEAGLEsViewEx2020, proctorWeakConnectionStellar2025}; a higher value indicates a greater contribution of ex-situ stars relative to in-situ star formation.

Fig.~\ref{fig:morphology_merger} shows the median galaxy properties at $z=0$ as a function of $\fex$ in bins of stellar mass.
Strikingly, most properties vary monotonically with $\fex$ at fixed stellar mass, and their simultaneous co-variation allows us to construct a plausible picture of merger-driven galaxy evolution.
Galaxies with higher $\fex$ are systematically larger, consistent with the virial theorem argument that galaxies can grow in size by merging with smaller systems \citep[see][]{coleHierarchicalGalaxyFormation2000, naabMinorMergersSize2009, hopkinsDiscriminatingPhysicalProcesses2010, wang26}.
$\rm D/T$ decreases monotonically with $\fex$, consistent with mergers destroying the ordered rotational support of progenitor discs \citep{coleHierarchicalGalaxyFormation2000, wang26}.
Triaxiality rises steeply with $\fex$, consistent with mergers imprinting random orbital configurations on the remnant's stellar distribution and naturally producing triaxial and prolate morphologies \citep{ebrovaGalaxiesProlateRotation2017, liOriginPropertiesMassive2018, lagosDiverseNatureFormation2022, wang26}.
Finally, $M_{\rm BH}/M_{\rm star}$ increases monotonically with $\fex$, consistent with the theoretical picture that merger-driven disc disruption and gas inflow into the galactic centre fuel SMBH accretion, in addition to the direct coalescence of black holes during the merger itself \citep{barnesTransformationsGalaxiesII1996, kauffmannUnifiedModelEvolution2000, bowerBreakingHierarchyGalaxy2006, malbonBlackHoleGrowth2007,dimatteoEnergyInputQuasars2005, daviesGalaxyMergersCan2022}.
What Fig.~\ref{fig:morphology_merger} uniquely demonstrates is that these effects are not independent: they are all facets of the same merger-driven process, operating simultaneously and leaving a coherent imprint across the full morphological and star formation properties of the galaxy population.

A notable feature of Fig.~\ref{fig:morphology_merger} is that the quiescent fraction and the sSFR are insensitive to $\fex$ at $\fex \lesssim 0.3$ in all stellar mass bins, and in the lowest mass bin they show no dependence on $\fex$ at all\footnote{The star formation status appears similarly insensitive to $\fex$ in the highest mass bin, but this is due to contamination by rejuvenated galaxies, which we explore further in \S\,\ref{sub:reju}.}.
This behaviour reveals that mergers do not directly lead to quenching, but act indirectly by promoting SMBH growth to the point where AGN feedback becomes sufficiently powerful to expel gas and suppress star formation \citep{bowerBreakingHierarchyGalaxy2006, crotonManyLivesActive2006, daviesGalaxyMergersCan2022}.
Only when the integrated merger activity is substantial, reflected in high $\fex$, does the accumulated SMBH mass reach the point where AGN feedback becomes effective, at which point the quiescent fraction rises steeply.
The connection between merger activity and quenching is therefore mediated through $M_{\rm BH}$, as reflected in the co-variation of $M_{\rm BH}/M_{\rm star}$ and $f_\mq$ across the full range of $\fex$ in Fig.~\ref{fig:morphology_merger}.
Together, these results establish a unified picture in which mergers simultaneously reshape galaxy structure, build up the central black hole, and ultimately drive quenching through AGN feedback.

So far we have quantified merger activity through $\fex$, which captures the cumulative stellar mass contributed by mergers over a galaxy's lifetime but not the properties of individual merger events.
Galaxies with similar $\fex$ and stellar mass may have assembled this mass through mergers with different mass ratios, which could plausibly contribute to the trends identified above.
In Appendix~\ref{sec:the_impact_of_galaxy_mergers} we test the role of merger mass ratio directly, by replacing $\fex$ with the stellar mass ratio of the single most massive merger along each galaxy's main branch.
The resulting trends are qualitatively similar but systematically weaker than those with $\fex$, demonstrating that the cumulative contribution of multiple mergers, rather than the properties of any single merger event, drives the structural transformation described here.

However, the merger-driven picture leaves an important aspect of the morphology--quenching connection unexplained.
Mergers grow galaxies in size: \citet{wang26} present the first statistical analysis of the impact of mergers on galaxy size in a cosmological hydrodynamical simulation, finding that a single event increases galaxy size by $\approx0.03$--$0.1$ dex, with repeated mergers compounding the effect.
A purely merger-driven quenching scenario would predict quiescent galaxies to be more extended than their star-forming counterparts at fixed stellar mass, which is the opposite of the results from simulations \citep[e.g.][]{furlongSizeEvolutionNormal2017, genelSizeEvolutionStarforming2018, ludlowEvolutionSizesAngular2026} and observations \citep[e.g.][]{vanderwel3DHSTCANDELSEvolution2014}.
This suggests an alternative origin for the compactness of quiescent galaxies.

Fig.~\ref{fig:morphology_quenching_fexsitu} shows the morphological properties of central galaxies at $z = 0$ as a function of stellar mass, split simultaneously by star formation status and ex-situ stellar mass fraction into three bins of $\fex \in [0.0, 0.2)$, $[0.2, 0.4)$, and $[0.4, 1.0)$.
This decomposition allows us to isolate the morphological differences between star-forming and quiescent galaxies at fixed merger history, separating the contribution of mergers from any additional internally-driven channel.
As expected from Fig.~\ref{fig:morphology_merger}, galaxies with higher $\fex$ are systematically larger, less disc-dominated, and more triaxial within both populations.
Crucially, however, at fixed stellar mass \textit{and} fixed $\fex$, quiescent galaxies remain systematically more compact, have lower $\rm D/T$, and higher triaxiality than their star-forming counterparts, especially in the lowest $\fex$ bin.
This residual morphological difference, present even after controlling for merger history, demonstrates that mergers alone cannot account for the morphology--quenching connection, and points toward an additional, internally-driven channel through which quiescent galaxies acquire their distinctive morphologies.

\subsection{Morphology-modulated SMBH growth}
\label{sub:morphology_modulated_smbh_growth} 

\begin{figure}
  \begin{center}
    \includegraphics[width=0.95\linewidth]{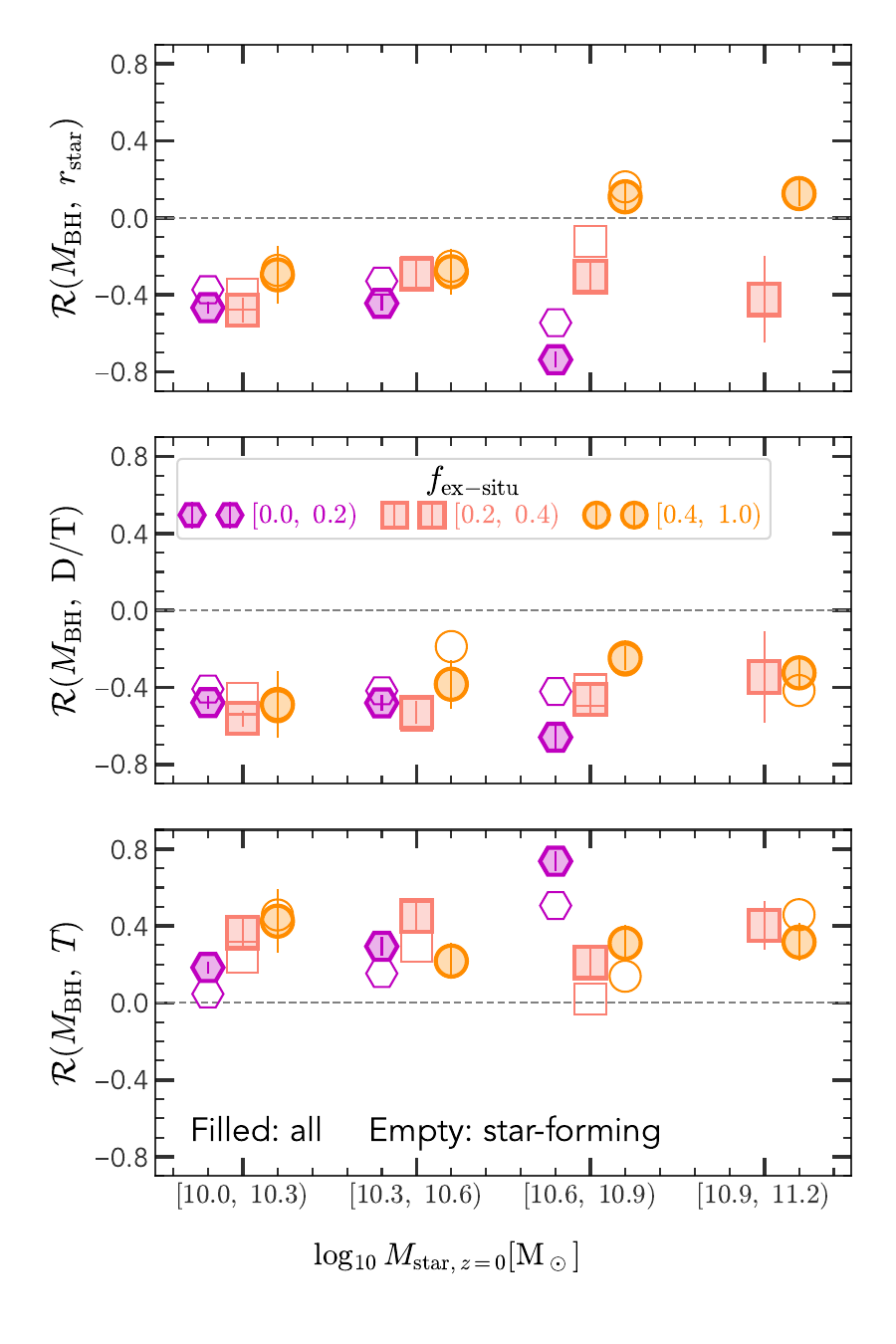}
  \end{center}
  \caption{
  Spearman's rank correlation coefficients between $M_{\rm BH}$ and galaxy morphological properties, $r_{\rm star}$ (top), $\rm D/T$ (middle), and triaxiality $T$ (bottom), measured at $z=0$ within bins of stellar mass and ex-situ stellar mass fraction (hexagons: $\fex\in [0.0,~0.2)$, squares: $\fex\in [0.2,~0.4)$, circles: $\fex\in [0.4,~1.0)$).
  Filled and open symbols show results for all galaxies and star-forming galaxies, respectively.
  The dashed line marks zero correlation.
  The correlation between $r_{\rm star}$ and $M_{\rm BH}$ changes sign with $\fex$: merger-poor galaxies show a moderate negative correlation, consistent with compact galaxies growing their SMBHs more efficiently, whereas merger-rich galaxies show a positive correlation, reflecting the concurrent merger-driven size growth and SMBH growth.
  $\rm D/T$ exhibits a moderate negative correlation ($\mathcal R\sim -0.5$) with $M_{\rm BH}$ across all bins, indicating that galaxies with less rotational support grow their SMBH more efficiently regardless of merger history.
  Triaxiality shows a weak to moderate positive correlation with $M_{\rm BH}$ across most bins.
  The comparable correlation strengths between star-forming and quiescent galaxies indicate that morphology-modulated SMBH growth is already operating during the star-forming phase.
  }
  \label{fig:correlation_to_mbh}
\end{figure}

The morphological differences between star-forming and quiescent galaxies identified above persist even after controlling for merger history, suggesting that an additional physical mechanism couples galaxy structure to SMBH growth and quenching independently of mergers.
A natural candidate is a direct connection between galaxy morphology and SMBH accretion: if the structural properties of a galaxy govern how readily gas can reach the central SMBH, then morphology itself may modulate SMBH growth and ultimately determine which galaxies are quenched \citep{hopkinsAnalyticModelAngular2011, shankarSizeEvolutionSpheroids2013, angles-alcazarGravitationalTorquedrivenBlack2017, robertsGalaxyDiscsRegulate2026}.
To test this hypothesis, we compute Spearman's rank correlation coefficient $\mathcal{R}$ between $M_\mathrm{BH}$ and each morphological property, $\rm D/T$, $r_\mathrm{star}$, and $T$, within bins of stellar mass and $\fex$, separately for star-forming and quiescent galaxies.
The results are shown in Fig.~\ref{fig:correlation_to_mbh}.

The correlation between $r_{\rm star}$ and $M_{\rm BH}$ reveals a qualitatively different behaviour depending on the value of $\fex$.
For merger-poor galaxies ($\fex < 0.2$), the anti-correlation indicates that more compact galaxies host more massive SMBHs at fixed stellar mass.
This is a key result: in the absence of significant merger activity, the compactness of a galaxy is correlated with its SMBH mass, as compact galaxies have higher central gas densities, which directly sustain higher SMBH accretion rates.
Such galaxies are therefore driven toward higher SMBH masses at which AGN feedback becomes capable of quenching star formation, naturally producing the size deficit of quiescent galaxies relative to their star-forming counterparts at fixed stellar mass and $\fex$ seen in Fig.~\ref{fig:morphology_quenching_fexsitu}.
For merger-rich galaxies ($\fex > 0.4$), the correlation becomes near zero or even positive, reflecting the fact that mergers simultaneously drive size growth and SMBH growth, overwhelming and reversing the intrinsic compactness--SMBH connection.

The D/T--$M_\mathrm{BH}$ correlation is negative across all stellar mass and $\fex$ bins, with a typical strength of $\mathcal{R} \sim -0.5$ \citep[see also][]{correaDependenceGalaxyStellartohalo2020}.
This indicates that galaxies with less rotational support grow their SMBHs more efficiently at fixed stellar mass and merger history.
Physically, a lower D/T implies a larger dispersion-supported stellar component with reduced ordered angular momentum, which is associated with a less axisymmetric gravitational potential that facilitates gas inflow toward the galactic centre and sustains higher SMBH accretion rates \citep[see also][]{hopkinsAnalyticModelAngular2011, angles-alcazarGravitationalTorquedrivenBlack2017}.
Crucially, this correlation is present and of comparable strength also in star-forming galaxies across all $\fex$ bins, demonstrating that morphology-modulated SMBH growth is not a consequence of quenching but rather an ongoing process already operating during the star-forming phase.

Finally, triaxiality shows a weak to moderate positive correlation with $M_\mathrm{BH}$ across most bins.
A more triaxial stellar distribution implies reduced ordered rotation and a less axisymmetric gravitational potential, both of which are conducive to non-circular gas orbits and enhanced inflow toward the centre \citep{hopkinsAnalyticModelAngular2011, dadoDynamicalDisequilibriumDwarf2026}.
Together, the correlations shown in Fig.~\ref{fig:correlation_to_mbh} establish that galaxy morphology modulates SMBH growth independently of merger activity, with compactness and the absence of rotational support acting as the primary structural drivers.
This morphology-modulated channel provides a coherent explanation for the residual morphological differences between star-forming and quiescent galaxies at fixed $\fex$.

\subsection{AGN-feedback-driven quenching facilitated by galaxy size}
\label{sub:agn_feedback_driven_quenching_facilitated_by_galaxy_size} 

\begin{figure}
  \begin{center}
    \includegraphics[width=0.95\linewidth]{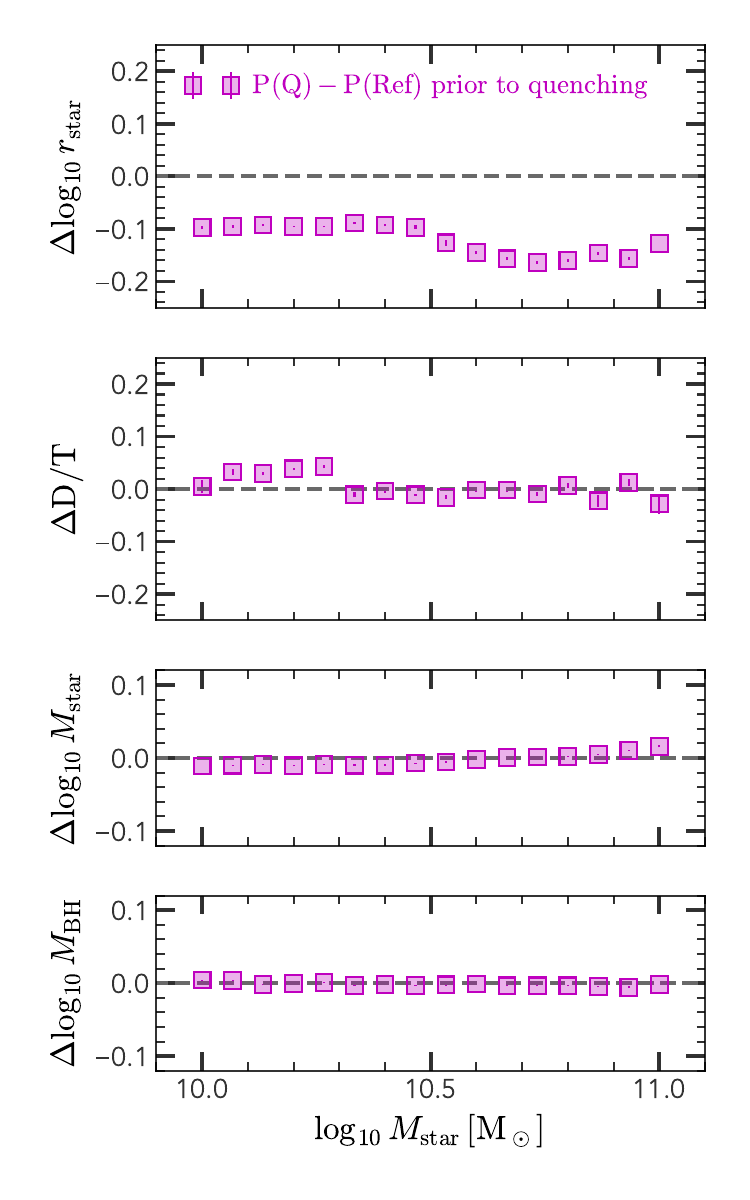}
  \end{center}
  \caption{
    Differences in morphological and structural properties between quiescent central galaxies at $z = 0$ (fiducial sample) and star-forming galaxies matched in stellar mass and SMBH mass at the most recent snapshot prior to quenching (reference sample).
    From top to bottom, the panels show the median differences in $\log_{10} r_\mathrm{star}$, $\rm D/T$, $\log_{10} M_\mathrm{star}$, and $\log_{10} M_\mathrm{BH}$, as a function of stellar mass in the snapshot prior to quenching.
    The dashed line marks zero difference.
    The stellar mass and SMBH mass panels confirm the matching is successful.
    Galaxies destined to quench are systematically more compact by $\sim 0.1$--0.2 dex than their star-forming counterparts at fixed stellar mass and SMBH mass, while their D/T values are indistinguishable, demonstrating that it is compactness rather than disc fraction that distinguishes galaxies that quench from those that remain star-forming at fixed black hole mass.
  }
  \label{fig:quenching_morphology_role}
\end{figure}

We have shown that galaxy morphology and SMBH growth are correlated.
Feedback from SMBHs has been shown to play an important role in quenching massive galaxies \citep{bluckHowCentralSatellite2020, piotrowskaQuenchingStarFormation2022, wangBlackHolesRegulate2024}.
However, it remains unclear whether galaxy morphology plays an additional role in facilitating AGN-feedback-driven quenching that is independent of its influence on SMBH growth.
\citet{jiangDissectingMassQuenching2025} addressed this question using the TNG50 simulation and found that galaxy size modulates the effectiveness of AGN feedback: the kinetic AGN feedback in the IllustrisTNG model can only directly suppress star formation within a limited central region ($\approx 1-2\,\rm kpc$), so that compact galaxies, whose stellar body lies largely within this region, are quenched rapidly and directly, while more extended galaxies can only be quenched preventively through the suppression of gas replenishment on longer timescales.

To test whether a similar size-dependent quenching mode operates in EAGLE, we construct a matched sample comparison.
For each quiescent central galaxy at $z = 0$ (the fiducial sample), we trace it back to the most recent snapshot at which it was still star-forming, and identify all star-forming galaxies at that snapshot with similar stellar mass (within 0.05 dex) and SMBH mass (within 0.1 dex), forming a reference sample.
We further require reference galaxies to lie on the main progenitor branch of $z = 0$ galaxies and to have $\mathrm{sSFR} > 10^{-10.5}\,\mathrm{yr}^{-1}$ at $z = 0$, ensuring they are genuinely star-forming descendants rather than galaxies approaching quiescence.
By construction, the two samples are matched in $M_\mathrm{star}$ and $M_\mathrm{BH}$ at the same redshift, as confirmed by the bottom two panels of Fig.~\ref{fig:quenching_morphology_role}.
Any remaining morphological differences therefore cannot be attributed to differences in stellar or black hole mass, nor to evolutionary stage.

Fig.~\ref{fig:quenching_morphology_role} shows the median differences in $\log_{10} r_\mathrm{star}$, D/T, $\log_{10} M_\mathrm{star}$, and $\log_{10} M_\mathrm{BH}$ between the fiducial and reference samples, evaluated at the same redshift snapshot prior to quenching.
Galaxies destined to quench are systematically more compact by $\sim 0.1$--$0.2$\,dex than their star-forming counterparts matched in $M_\mathrm{star}$ and $M_\mathrm{BH}$, across the full range of present-day stellar mass.
By contrast, the difference in $\mathrm{D/T}$ is negligible.
These results indicate that, at fixed black hole mass, it is the compactness of a galaxy rather than the absence of rotational support that determines whether AGN feedback is sufficient to drive quenching.
This is consistent with the TNG50 results of \citet{jiangDissectingMassQuenching2025}, despite these two simulations having distinct AGN feedback implementations.

\subsection{Post-quenching morphological evolution}
\label{sub:post_quenching_morphology_evolution} 

\begin{figure*}
  \begin{center}
    \includegraphics[width=0.9\linewidth]{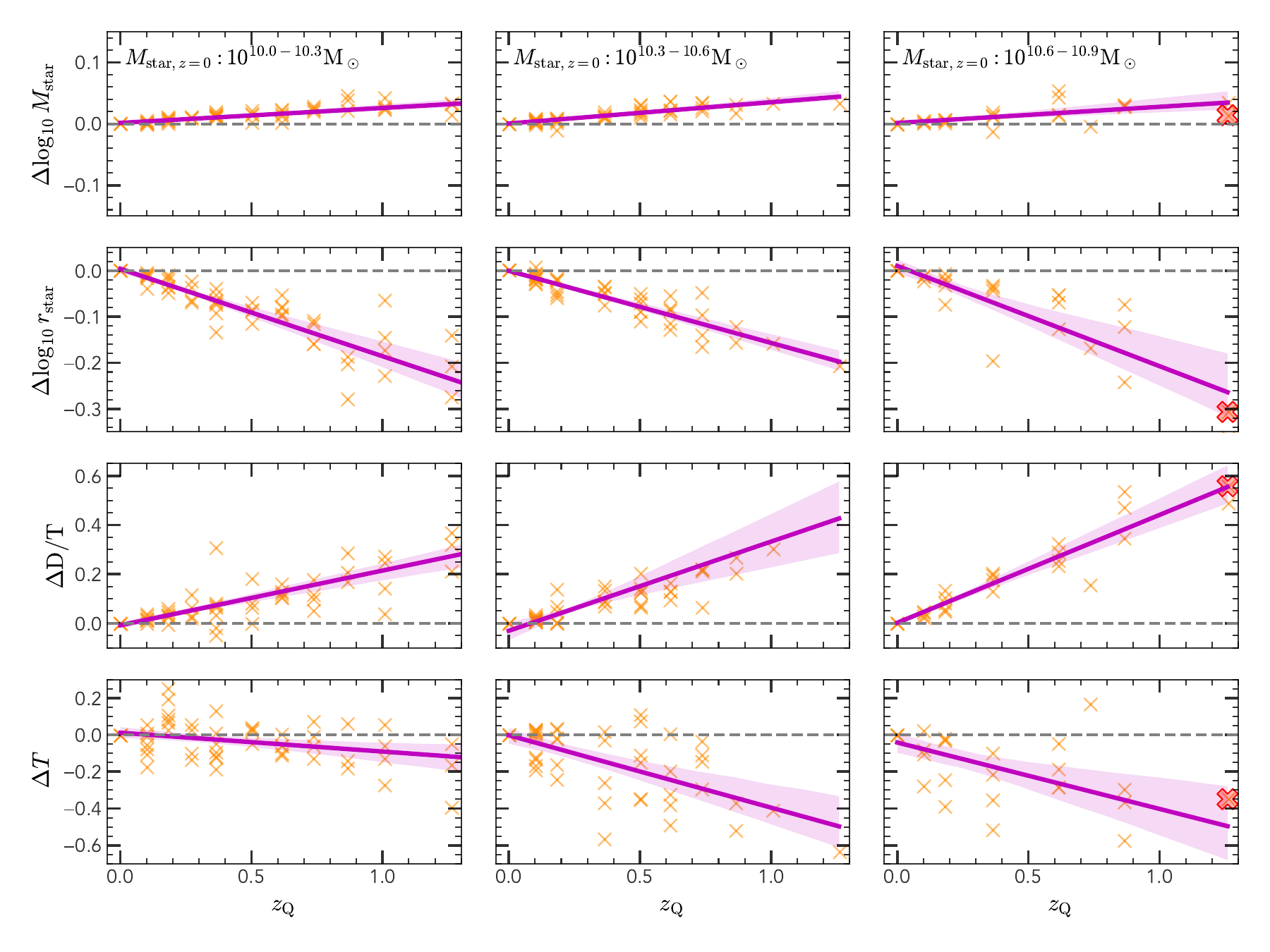}
  \end{center}
  \caption{
  Morphological evolution after quenching for quiescent central galaxies with $f_{\rm ex-situ} < 0.2$.
  For each galaxy, $z_\mq$ is identified as the latest snapshot at which the galaxy becomes quiescent.
  Each panel shows $\Delta P\equiv P(z_\mq) - P(z=0)$ for, from top to bottom, $\log_{10}M_{\rm star}/{\rm M_\odot}$, $\log_{10} r_{\rm star}$, $\rm D/T$, and $T$.
  Orange crosses show individual galaxies; the magenta line and the shaded region show the linear best fit and its $1\sigma$ uncertainty.
  The big red cross marks the example galaxy shown in Fig.~\ref{fig:example1}.
  Despite negligible stellar mass growth after quenching, galaxies grow significantly in size, lose their disc structure, and increase in triaxiality, with the magnitude of these changes increasing with $z_\mq$.
  }
  \label{fig:morphology_evolution_after_quenching}
\end{figure*}

The morphology--$M_{\rm BH}$ correlations established in \S\,~\ref{sub:morphology_modulated_smbh_growth} identify compactness and low rotational support as the structural conditions that promote SMBH growth and ultimately drive quenching in merger-poor galaxies.
A natural question follows: does morphological transformation cease once star formation is quenched, or does it continue to evolve under the influence of internal secular processes?
To address this, we trace the structural evolution of merger-poor ($\fex < 0.2$) quiescent central galaxies between the epoch of quenching, $z_\mathrm{Q}$, and $z = 0$.
For each galaxy, $z_\mathrm{Q}$ is defined as the latest snapshot at which their SFR crosses the quenching threshold.
Fig.~\ref{fig:morphology_evolution_after_quenching} shows the change $\Delta P \equiv P(z_\mathrm{Q}) - P(z=0)$ in each structural property as a function of $z_\mathrm{Q}$, for three bins of present-day stellar mass, where $P$ refers to $\log_{10}M_{\rm star}$, $\log_{10}r_{\rm star}$, $\rm D/T$, and  $T$ from top to bottom.

The stellar mass panel shows that $\Delta \log_{10} M_\mathrm{star}$ is consistent with zero or slightly positive across the full range of $z_\mathrm{Q}$, indicating that these galaxies undergo negligible stellar mass growth after quenching\footnote{The marginal positive offset reflects the gradual stellar mass loss through stellar winds after star formation}.
Any subsequent structural evolution must therefore be driven by internal dynamical processes.
Despite this, the remaining panels reveal substantial morphological transformation continuing after quenching, with the magnitude of all changes increasing systematically with $z_\mq$: the longer a galaxy has been quenched, the more its morphology has evolved since star formation ceased.
Generally, galaxies that quenched at $z_\mq \approx 1$ have grown in size by $\approx 0.2 \,\rm dex$, decreased their $\rm D/T$ by $\approx 0.2-0.4$, and increased their triaxiality by $\approx 0.1-0.4$, with the precise values depending on the present-day stellar mass.
This continued size growth after quenching acts to gradually erode the size deficit established at the moment of quenching, and is consistent with the picture seen in Fig.~\ref{fig:evolution_afo_mstar2}: the size difference between star-forming and quiescent galaxies is larger at higher redshift and diminishes toward $z=0$ as quiescent galaxies continue to grow.
A similar post-quenching size growth at approximately constant stellar mass has been reported in IllustrisTNG by \citet{genelSizeEvolutionStarforming2018}.

Together, these results reveal that the morphology--quenching connection arises through multiple physically distinct channels.
First, merger activity drives a coherent co-evolution of galaxy structure, SMBH growth, and star formation quenching, simultaneously making galaxies more extended, more triaxial, and less disc-dominated while building up the central black hole to masses at which AGN feedback becomes capable of quenching the galaxy.
Second, in the absence of significant merger activity, it is the structural properties of the galaxy itself, in particular its compactness and lack of rotational support, that modulate SMBH growth and determine which galaxies are preferentially quenched.
Third, at fixed SMBH mass, galaxies destined to quench are systematically more compact than their star-forming counterparts, suggesting that galaxy size plays an additional role in facilitating AGN-feedback-driven quenching independently of black hole mass.
Fourth, morphological transformation does not cease at the moment of quenching but continues to operate through internal secular processes long after star formation has been suppressed.

\section{Prolate galaxies formed out of disc instability}
\label{sec:prolate_galaxies_formed_out_of_disc_instability}

Fig.~\ref{fig:morphology_quenching_fexsitu} reveals a striking anomaly among quiescent galaxies with $10^{10.6} \leq M_{{\rm star}, z=0} < 10^{10.9}\,\msun$ and $\fex < 0.2$: their median triaxiality is unusually high, comparable to that of merger-rich quiescent galaxies at the same stellar mass, despite these galaxies having experienced minimal merger activity throughout their lifetime.
What physical process is responsible for building such prolate morphologies in the absence of mergers?
In this section we investigate the origin of this anomaly by tracing the evolutionary histories of these galaxies, examining individual examples, and characterising the resulting shape distribution and demographics of the prolate population.

\subsection{A merger-free pathway to morphological transformation}
\label{sub:a_merger_free_pathway_to_morphological_transformation} 

\begin{figure}
  \begin{center}
    \includegraphics[width=0.95\linewidth]{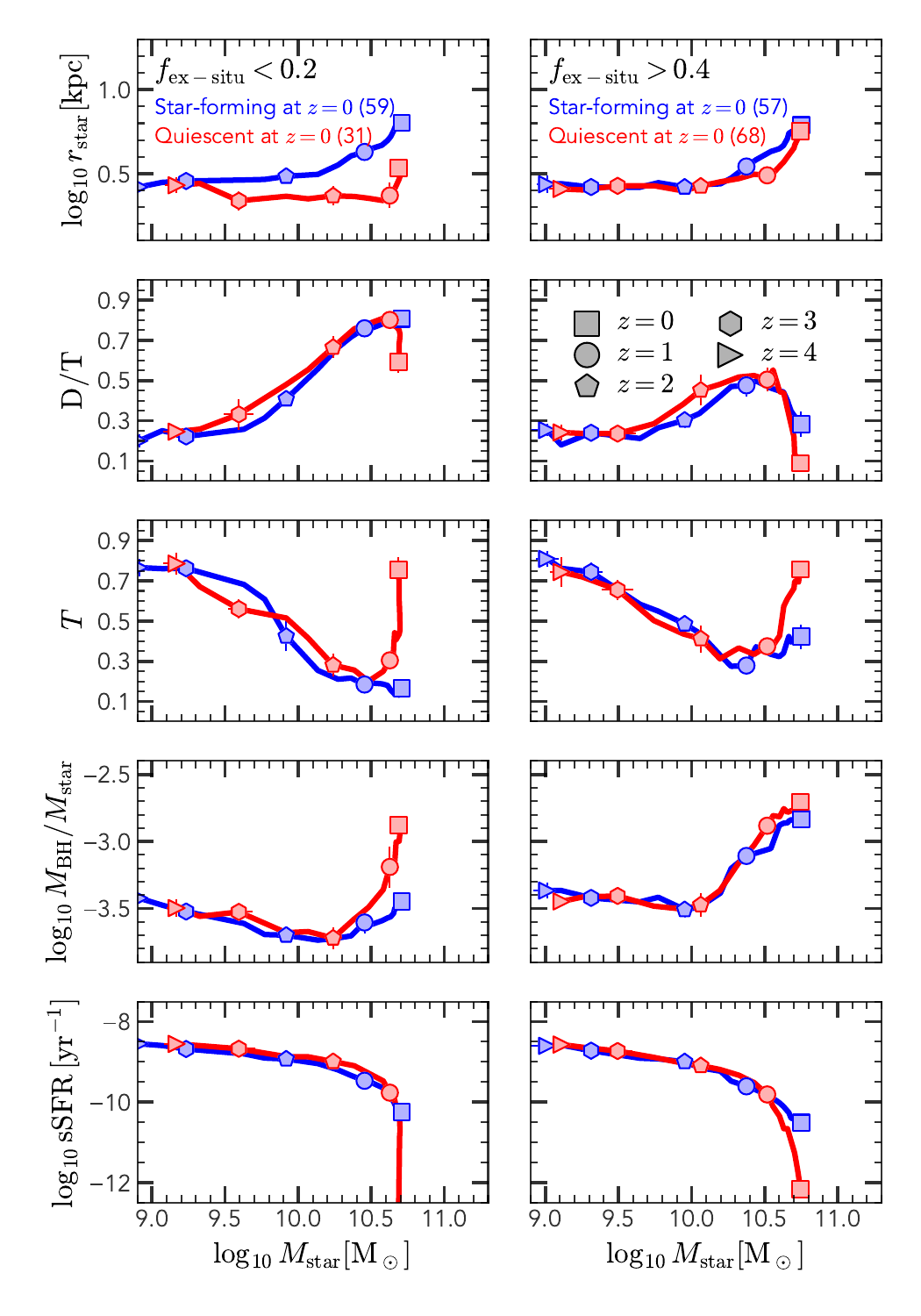}
  \end{center}
  \caption{
    Progenitor trajectories for star-forming and quiescent central galaxies with $10.6\leq \log_{10}(M_{\rm star, z=0}/{\rm M_\odot})< 10.9$, split into the $f_{\rm ex-situ} < 0.2$ (left) and $f_{\rm ex-situ} > 0.4$ (right) subsamples, where $f_{\rm ex-situ}$ is the present-day \textit{ex-situ} stellar mass fraction.
    From top to bottom, the rows show the evolution of galaxy size ($r_{\rm star}$), disc-to-total stellar mass ratio ($\rm D/T$), triaxiality ($T$), BH-to-stellar mass ratio ($M_{\rm BH}/M_{\rm star}$), and sSFR along the main progenitor branch.
    Different markers represent different redshift snapshots from $z=0$ to $z=4$.
    Numbers in parentheses give the size of each subsample.
    At high $f_{\rm ex-situ}$ the two populations follow similar trends in size, $\rm D/T$, $T$ and $M_{\rm BH}/M_{\rm star}$ throughout their growth, but reach different $z=0$ values, with the galaxies that quench ending at lower $\rm D/T$, higher $T$ and higher $M_{\rm BH}/M_{\rm star}$.
    At low $f_{\rm ex-situ}$ the galaxies that quench are more compact than their star-forming counterparts by $\approx0.3$ dex over most of their growth, and they undergo a near-vertical transformation at almost constant stellar mass in which $\rm D/T$ falls and $T$ rises sharply.
  }
  \label{fig:evolution_fex_subsample}
\end{figure}

\begin{figure*}
  \begin{center}
    \includegraphics[width=0.95\linewidth]{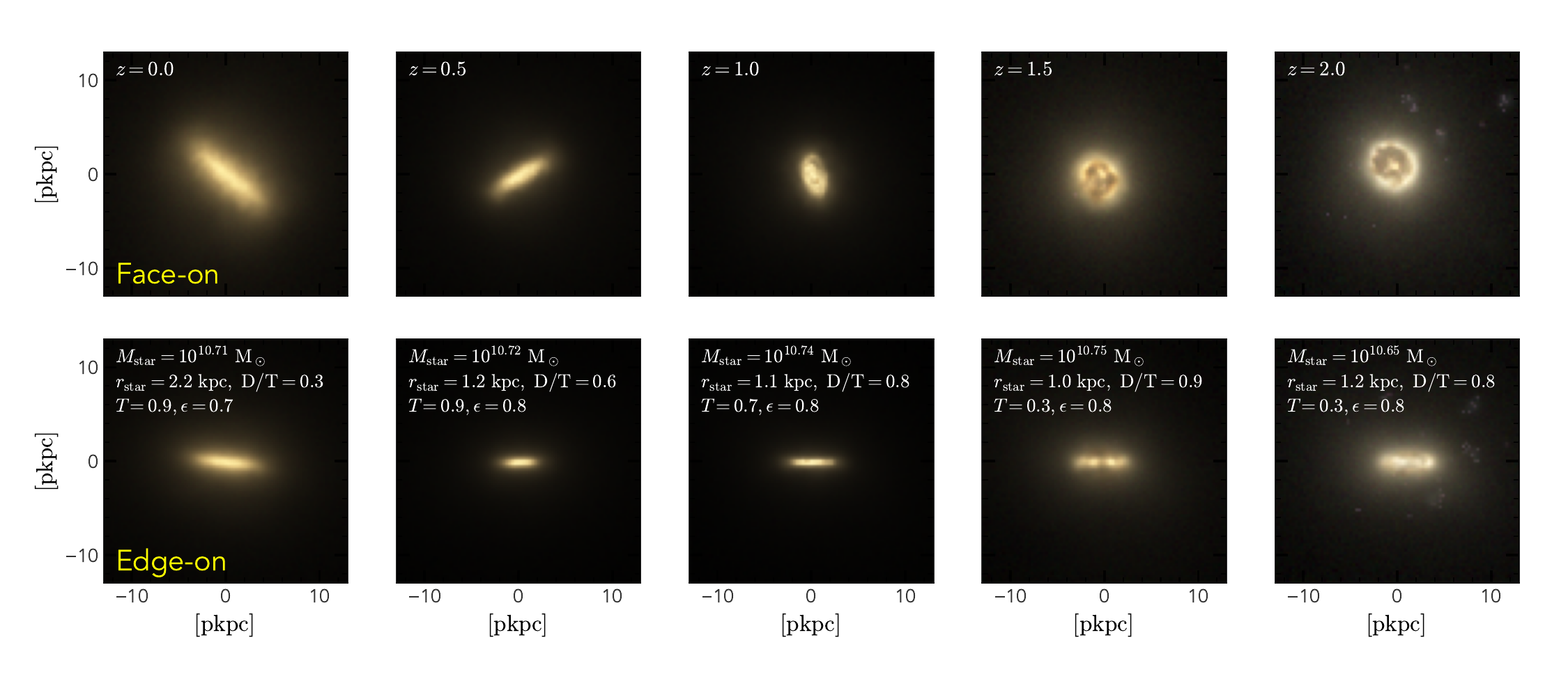}
  \end{center}
  \caption{
    Mock images of an example quiescent central galaxy undergoing morphological transformation driven by internal disc instability, shown from $z=2$ (right) to $z=0$ (left).
    The top and bottom rows show face-on and edge-on projections, respectively, each spanning $25\times 25$ pkpc.
    Structural parameters measured at each snapshot are annotated in lower panels.
    The galaxy undergoes negligible stellar mass growth since $z\approx 1.5$, yet experiences dramatic structural transformation from an initial compact disc ($\rm D/T \approx 0.9$, $T\approx 0.3$) at $z= 1.5$ to a prominent bar-like prolate galaxy at $z= 0$ ($\rm D/T\approx 0.3$, $T\approx 0.9$), accompanied by an increase in $r_{\rm star}$ from 1.0 pkpc to 2.2 pkpc.
    This example illustrates the internally-driven pathway identified in Fig.\,\ref{fig:evolution_fex_subsample}, in which disc instability induces bar formation.
  }
  \label{fig:example1}
\end{figure*}

\begin{figure}
  \begin{center}
    \includegraphics[width=0.95\linewidth]{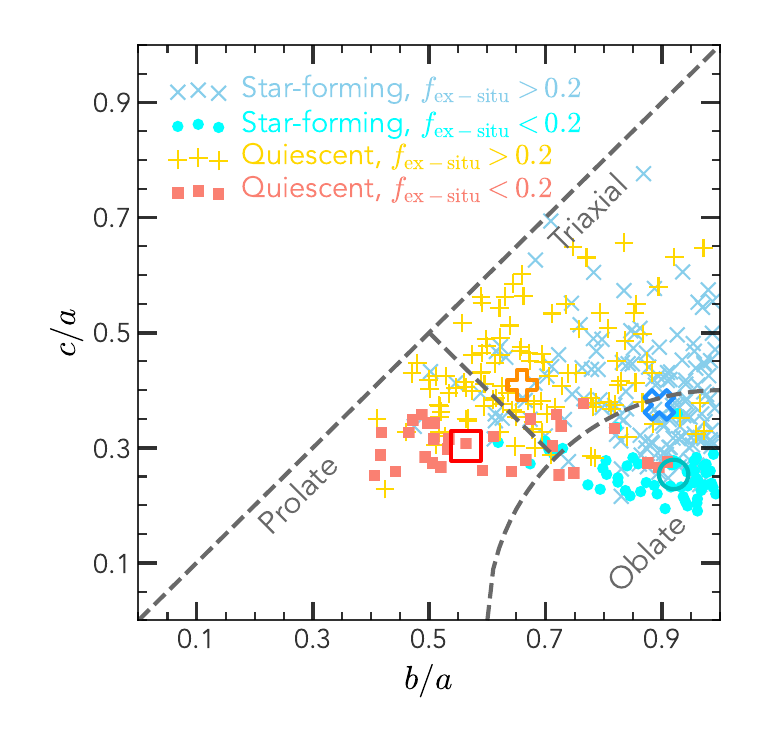}
  \end{center}
  \caption{
    Intrinsic shape distribution of central galaxies with $10.6\leq \log_{10}(M_{\rm star, z=0}/{\rm M_\odot})< 10.9$ in the $b/a-c/a$ plane, where $a > b>c$ are the principal axes of the stellar mass distribution.
    The dashed lines divide the plane into prolate ($c\approx b \ll a$), oblate ($c \ll b \approx a$), and triaxial regimes.
    Galaxies are split by star formation status and \textit{ex-situ} fraction: star-forming with $f_{\rm ex-situ} > 0.2$, star-forming with $f_{\rm ex-situ} < 0.2$, quiescent with $f_{\rm ex-situ} > 0.2$, and quiescent with $f_{\rm ex-situ} < 0.2$.
    Large symbols mark the median of each subsample.
    Low-$f_{\rm ex-situ}$ galaxies shift almost purely horizontally from the oblate region, where they are star-forming, to the prolate region, where they become quiescent, with little change in $c/a$, consistent with a transformation from a compact disc to an elongated prolate structure at roughly constant thickness.
    High-$f_{\rm ex-situ}$ galaxies preferentially occupy the triaxial region, with star-forming galaxies lying closer to the oblate boundary and quiescent galaxies displaced toward more prolate configurations.
  }
  \label{fig:prolate_distribution.pdf}
\end{figure}

To understand the origin of this anomaly, we trace the progenitors of merger-poor ($\fex < 0.2$) and merger-rich ($\fex > 0.4$) subsamples separately, for central galaxies with $10^{10.6} \leqslant M_{{\rm star},z=0} < 10^{10.9}\,\msun$.
Fig.~\ref{fig:evolution_fex_subsample} shows the median evolutionary trajectories of $r_{\rm star}$, D/T, triaxiality $T$, and $M_{\rm BH}/M_{\rm star}$ along the main progenitor branch, separately for star-forming and quiescent descendants in each $\fex$ subsample.

Among high-$\fex$ galaxies, the star-forming and quiescent descendants follow similar trends in size, D/T, triaxiality and SMBH mass as they accumulate stellar mass, and their sizes converge at $z=0$.
The two populations separate in their final values, with the quiescent descendants ending at lower D/T, higher triaxiality and higher $M_{\rm BH}/M_{\rm star}$.
The excess in $M_{\rm BH}/M_{\rm star}$ builds gradually from $M_{\rm star}\approx10^{10}\,\msun$ onwards, so merger-driven structural evolution leads to quenching only once the SMBH has grown massive enough for AGN feedback to become effective.

Among low-$\fex$ galaxies, the star-forming and quiescent descendants follow qualitatively \textit{distinct} trajectories, differing in two respects.
First, the size difference is present at all progenitor masses and widens with time: the quiescent subsample stays close to $\log_{10}(r_{\rm star}/{\rm kpc})\approx0.4$ while the star-forming subsample grows steadily in size, so the two end at $z=0$ separated by $\approx0.3$ dex.
Second, the quiescent subsample builds disc structure along the same track as the star-forming subsample, matching it in both D/T and triaxiality until $z\approx1$, then undergoes a near-vertical transformation at approximately constant stellar mass in which D/T declines steeply and triaxiality rises sharply.
The SMBHs of low-$\fex$ quiescent galaxies grow faster per unit of stellar mass growth than those of their high-$\fex$ counterparts, seen as a steeper slope in the $M_{\rm BH}/M_{\rm star}$--$M_{\rm star}$ panel.

The near-vertical evolutionary track of the merger-poor quiescent subsample raises the question of which physical process drives such a dramatic structural transformation at constant stellar mass.
To gain insight, we examine the morphological evolution of an individual galaxy drawn from this population.
Fig.~\ref{fig:example1} shows mock face-on and edge-on images of an example quiescent central galaxy with $\fex < 0.2$, traced from $z=2$ to $z=0$ \citep[see][]{trayfordOpticalColoursSpectral2017}.
At $z\gtrsim 1.5$, the galaxy is a compact, rotationally supported disc with ${\rm D/T} \approx 0.8$ and $T\approx 0.3$.
Despite negligible stellar mass growth thereafter, it undergoes a dramatic transformation: by $z=0$ it has developed a prominent elongated bar-like structure with ${\rm D/T}\approx 0.3$ and $T\approx 0.9$, while its effective radius has grown from 1.0 to 2.2 pkpc.
The face-on images reveal the progressive development of the bar from the initially compact disc, while the edge-on projections confirm that the transformation occurs at roughly constant thickness.
This example illustrates concretely the two-stage evolutionary track identified in Fig.~\ref{fig:evolution_fex_subsample}, and identifies internal disc instability leading to bar formation as the physical mechanism responsible.

The bar formation process illustrated in Fig.~\ref{fig:example1} leaves a distinct imprint on the three-dimensional shape of the stellar body that can be used to identify disc-instability-driven galaxies at the population level.
Fig.~\ref{fig:prolate_distribution.pdf} shows the intrinsic shape distribution of low- and high-$\fex$ galaxies in the $b/a - c/a$ plane, split by star formation status.
Low-$\fex$ galaxies undergo an almost purely horizontal displacement in the shape plane as they transition from star-forming to quiescent: their median shifts from the oblate region to the prolate region with little change in $c/a$, consistent with a transformation from a compact disc to an elongated bar-like structure at roughly constant thickness.
This horizontal shift is the geometric signature of bar formation, in which the disc elongates along one axis while its vertical extent remains largely unchanged, precisely the morphological transformation seen in the edge-on projections of Fig.~\ref{fig:example1}.
High-$\fex$ galaxies, by contrast, preferentially occupy the triaxial region, reflecting the random orbital configurations of material accreted in mergers that more often produce triaxial rather than purely prolate remnants.
The $c/a - b/a$ diagram therefore provides a clean diagnostic that distinguishes the disc-instability-driven channel from the merger-driven channel, with the former producing nearly purely prolate morphologies and the latter producing more generally triaxial configurations.

\subsection{Morphologcial transformation due to disc instability}
\label{sub:disc_instability_as_the_driving_mechanism} 

\begin{figure}
  \begin{center}
    \includegraphics[width=0.95\linewidth]{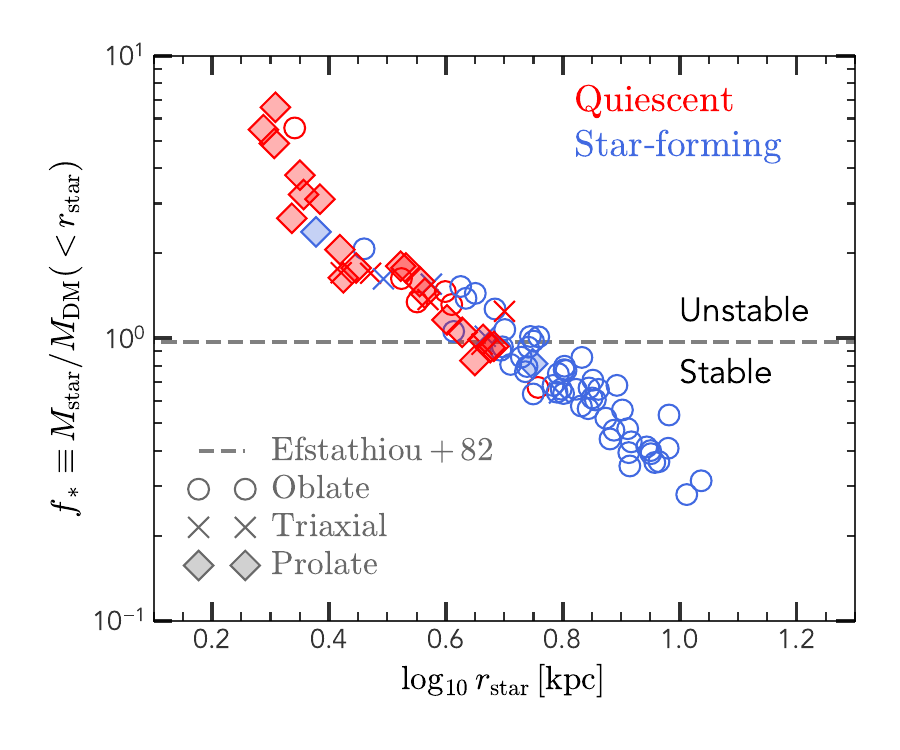}
  \end{center}
  \caption{
  Stellar-to-dark-matter mass fraction within the stellar half-mass radius, $f_{\rm star}$, as a function of galaxy size $r_{\rm star}$, for central galaxies with $\log_{10}(M_{\rm star}/{\rm M_\odot})\in [10.6,~10.9)$ and $f_{\rm ex-situ} < 0.2$.
  Red and blue symbols indicate quiescent and star-forming galaxies, respectively, while circles, pentagons, and plus signs denote oblate, triaxial, and prolate morphologies.
  The dashed line shows the threshold below which the disc is unstable according to \citet{efstathiouStabilityMassesDisc1982}.
  Quiescent galaxies are systematically more compact and more stellar-mass dominated than their star-forming counterparts, placing them preferentially inside the instability region.
  Prolate galaxies at $z=0$ similarly occupy the compact, baryon-dominated regime, consistent with their morphological transformation having been driven by disc instability caused by their high stellar mass concentration.
  }
  \label{fig:fstar}
\end{figure}

The mock images in Fig.~\ref{fig:example1} suggests internal disc instability as a plausible mechanism responsible for the dramatic rise in triaxiality at approximately constant stellar mass seen in the evolutionary tracks of merger-poor quiescent galaxies.
We now test whether the structural properties of these galaxies are indeed consistent with the conditions required for disc instability to occur.
The relevant criterion is that of \citet{efstathiouStabilityMassesDisc1982}, which predicts that a disc becomes unstable when its stellar mass sufficiently dominates over the dark matter within the disc radius, such that self-gravity overwhelms the stabilising influence of the dark matter halo.

Fig.~\ref{fig:fstar} tests this criterion directly by showing the stellar-to-dark-matter mass fraction within the stellar half-mass radius, $f_\mathrm{star} \equiv M_\mathrm{star}/M_\mathrm{DM}(<r_\mathrm{star})$, as a function of $r_\mathrm{star}$, for low-$\fex$ central galaxies with $10.6\leq \log_{10}(M_{\rm star, z=0}/{\rm M_\odot})< 10.9$.
Quiescent galaxies are systematically more compact and more baryon-dominated than their star-forming counterparts at the same stellar mass, placing them preferentially above the \citet{efstathiouStabilityMassesDisc1982} instability threshold.
Crucially, prolate galaxies at $z = 0$ occupy the same compact, baryon-dominated regime, consistent with their present-day morphology being a direct consequence of disc instability triggered by their high central stellar mass concentration.
Star-forming galaxies, by contrast, are more extended and dark-matter dominated within $r_\mathrm{star}$, placing them below the instability threshold and explaining why they retain their disc-dominated, oblate configurations.

Together, Figs.~\ref{fig:example1} and \ref{fig:fstar} establish a coherent physical picture: compact discs in which the stellar mass dominates the gravitational potential within the disc radius are susceptible to bar-forming instabilities that drive prolate morphologies and reduce rotational support, all without requiring any external perturbation.
This disc instability channel operates preferentially in the compact, merger-poor population, providing a natural explanation for the anomalously high triaxiality of low-$\fex$ quiescent galaxies identified in Section~\ref{sub:a_merger_free_pathway_to_morphological_transformation}.

\subsection{Demographics of the prolate galaxy population}
\label{sub:demographics_of_the_prolate_galaxy_population} 

\begin{figure}
  \begin{center}
    \includegraphics[width=0.95\linewidth]{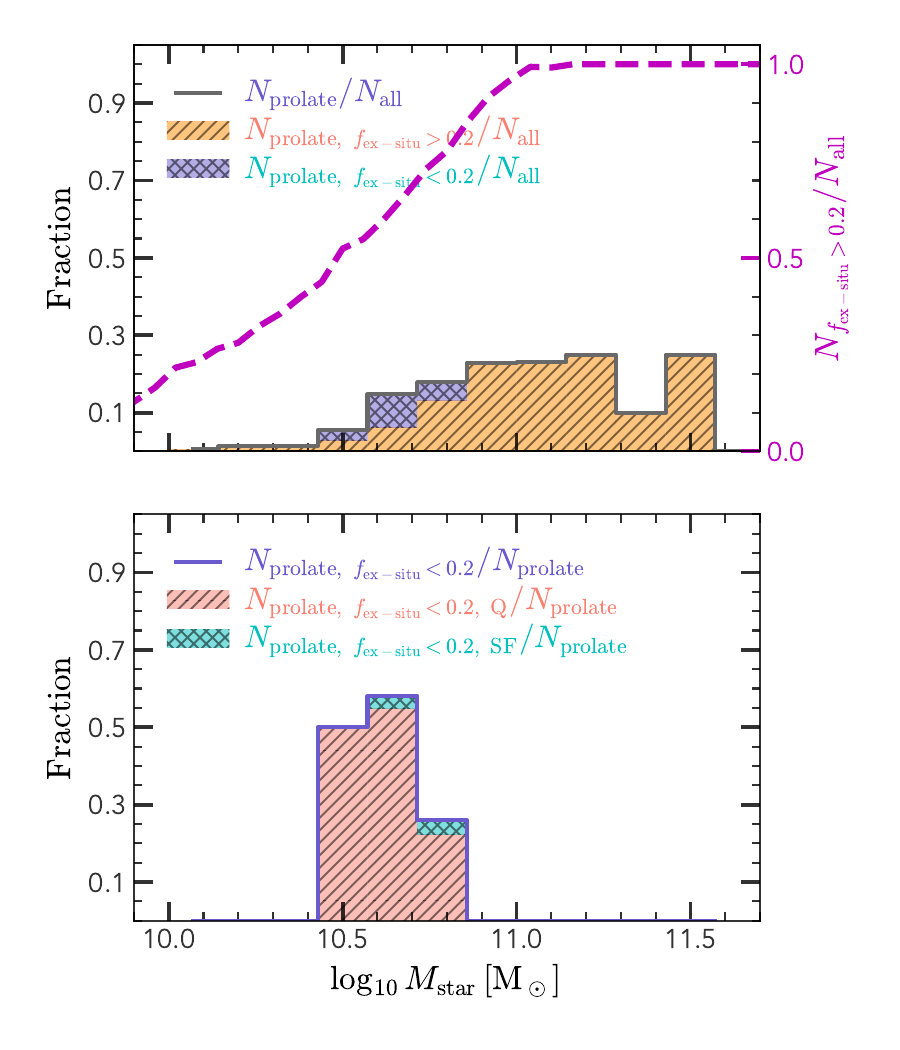}
  \end{center}
  \caption{
    Demographics of prolate galaxies as a function of stellar mass for central galaxies at $z=0$.
    The upper panel shows the fraction of all central galaxies that are prolate, decomposed into contributions from galaxies with $f_{\rm ex-situ} > 0.2$ and $f_{\rm ex-situ} < 0.2$.
    The magenta line shows the overall fraction of galaxies with $f_{\rm ex-situ}> 0.2$ for reference.
    The lower panel focuses on the low-$f_{\rm ex-situ}$ population, showing the fraction of all prolate galaxies that have $f_{\rm ex-situ} < 0.2$, further split into quiescent and star-forming subsamples.
    Around $10^{10.6}\,\rm M_\odot$, nearly half of all prolate galaxies have $f_{\rm ex-situ} < 0.2$, indicating that secular disc instability contributes significantly to prolate morphology at intermediate masses.
    This secular contribution diminishes towards higher masses, where the majority of galaxies are merger-rich ($f_{\rm ex-situ} > 0.2$) and mergers become the dominant pathway to prolate morphology, and towards lower masses, where discs are insufficiently massive to trigger bar instability.
  }
  \label{fig:prolate_demographic}
\end{figure}

Having established the two formation channels of prolate galaxies, we now quantify their relative demographic importance as a function of stellar mass.
Fig.~\ref{fig:prolate_demographic} shows the fraction of prolate galaxies at $z=0$, classified as in Fig.~\ref{fig:prolate_distribution.pdf}, split into merger-rich ($\fex > 0.2$) and merger-poor ($\fex < 0.2$) galaxies.
The overall prolate fraction rises from low values at $\log_{10}(M_{\rm star}/\msun) \approx 10$ to a broad peak at intermediate masses before declining at the highest masses, reflecting the interplay between the two formation channels across the mass function.
The merger-driven channel dominates at high stellar masses, where the majority of galaxies are merger-rich ($\fex > 0.2$) and the prolate fraction tracks the overall merger fraction closely.
At intermediate masses around $10^{10.6}\,\msun$, however, the low-$\fex$ contribution becomes substantial: nearly half of all prolate galaxies at this mass have $\fex < 0.2$, and the majority of these are quiescent, confirming that secular disc instability is a significant contributor to prolate morphology in this mass regime.
The secular contribution diminishes toward lower masses, where discs are insufficiently massive and baryon-dominated to satisfy the \citet{efstathiouStabilityMassesDisc1982} instability criterion, and toward higher masses, where AGN feedback suppresses in-situ star formation \citep{bowerBreakingHierarchyGalaxy2006}, so that further mass growth proceeds through mergers and the secular channel is closed off.
The lower panel further reveals that a non-negligible fraction of low-$\fex$ prolate galaxies remain star-forming, suggesting that bar formation can precede quenching in some cases, consistent with the picture of morphology-modulated SMBH growth discussed in Section~\ref{sub:morphology_modulated_smbh_growth}.

Together, these results demonstrate that the prolate galaxy population at $z = 0$ is not a monolithic outcome of a single formation channel, but rather the superposition of two physically distinct pathways: a merger-driven channel that dominates at high masses and produces triaxial remnants through the random deposition of accreted stellar material, and a secular disc instability channel that operates preferentially at intermediate masses and produces nearly purely prolate morphologies through bar formation in compact, baryon-dominated discs.

\section{Discussion}
\label{sec:discussion}

\subsection{The role of progenitor bias}
\label{sub:Progenitor bias} 

\begin{figure}
  \begin{center}
    \includegraphics[width=0.95\linewidth]{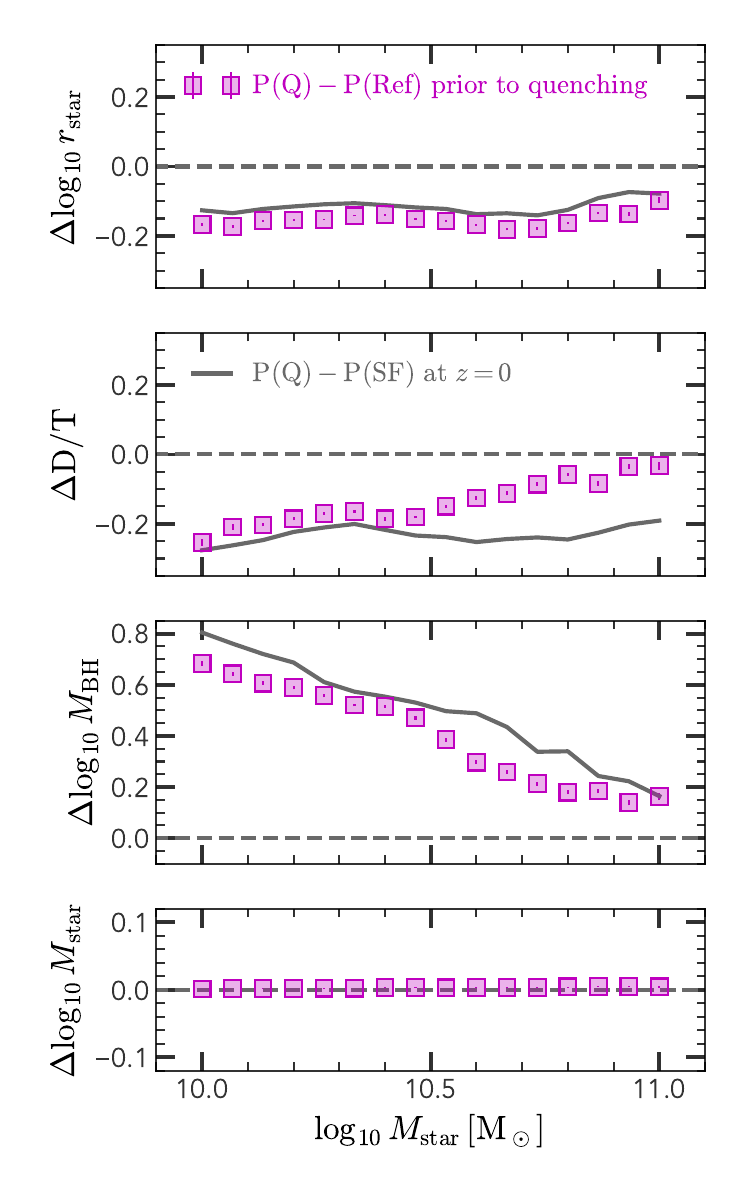}
  \end{center}
  \caption{
    Difference in structural properties between quiescent central galaxies and a reference star-forming sample matched in stellar mass, evaluated at the snapshot immediately prior to quenching (magenta squares; see \S\,\ref{sub:Progenitor bias} for details).
    The grey line shows the present-day difference between all quiescent and star-forming central galaxies at $z = 0$ for reference.
    From top to bottom, the panels show differences in $\log_{10} r_{\rm star}$, D/T, $\log_{10} M_{\rm BH}$, and $\log_{10} M_{\rm star}$.
    The stellar mass panel confirms the matching is successful.
    Galaxies destined to quench are already more compact, less disc-dominated, and host more massive SMBHs than their star-forming counterparts of the same mass immediately prior to quenching, confirming that these structural differences reflect genuine physical conditions rather than progenitor bias.
    The residual offset between the two epochs, a larger size deficit and a smaller D/T deficit prior to quenching relative to $z = 0$, is naturally explained by post-quenching secular evolution in which disc instability drives size growth and suppresses rotational support after star formation has ceased (see \S\,\ref{sub:post_quenching_morphology_evolution} for details).
  }
  \label{fig:progenitor_bias.pdf}
\end{figure}

A well-known alternative explanation for the structural differences between star-forming and quiescent galaxies is progenitor bias \citep{shankarAgeDependenceSizestellar2009, shankarAVOIDINGPROGENITORBIAS2015, lillySURFACEDENSITYEFFECTS2016}.
Under this hypothesis, quiescent galaxies are assumed to preserve the structural properties they possessed at the moment of quenching, so that the observed differences between the two populations at any given epoch arise naturally from the redshift evolution of galaxy scaling relations rather than from any physical process directly associated with quenching itself.
This idea has been invoked to explain both the size deficit \citep{lillySURFACEDENSITYEFFECTS2016} and the morphological differences \citep{tacchellaMorphologyStarFormation2019} of quiescent galaxies relative to their star-forming counterparts.

We test this hypothesis directly using the simulation by tracing each quiescent central galaxy back to the snapshot immediately prior to quenching, and identifying a reference sample of star-forming galaxies matched in stellar mass (within 0.05 dex) at the same snapshot that remain star-forming to $z = 0$.
The difference in structural properties between the quiescent and reference samples at this epoch is shown in Fig.~\ref{fig:progenitor_bias.pdf}, alongside the present-day difference between all quiescent and star-forming central galaxies at $z = 0$ for comparison.

Galaxies destined to quench already possess smaller sizes, lower D/T, and more massive SMBHs than their star-forming counterparts of the same stellar mass immediately prior to quenching.
At $M_{\rm star} \lesssim 10^{10.5}\,\msun$, the differences established prior to quenching are comparable to those observed at $z = 0$, indicating that little additional structural evolution occurs after quenching in this mass regime.
At higher masses, however, the size deficit prior to quenching is larger than at $z = 0$, while the D/T deficit is smaller.
This discrepancy is naturally accounted for by the post-quenching secular evolution identified in \S\,\ref{sub:post_quenching_morphology_evolution}: disc instability drives quiescent galaxies toward larger sizes and lower D/T after star formation has ceased, gradually eroding the size deficit and amplifying the D/T deficit toward their present-day values.
This immediately rules out progenitor bias as the dominant explanation: the structural differences are already in place before quenching occurs, and must therefore reflect genuine physical conditions that facilitate quenching rather than being a consequence of comparing populations assembled at different cosmic epochs.

Observations of post-starburst galaxies also disfavour the progenitor bias explanation.
\citet{almainiMassivePoststarburstGalaxies2017} found that post-starburst galaxies, which have quenched recently and still carry the spectral signature of a truncated burst of star formation, are more compact than star-forming galaxies of the same stellar mass at the same epoch \citep[see also][]{maltbyStructurePoststarburstGalaxies2018, chenSizeMassRelationPoststarburst2022, chengPoststarburstGalaxiesSDSSIV2024}.
These measurements contradict the progenitor bias explanation directly, since its core assumption is that galaxies have the same structure as the contemporaneous star-forming population immediately prior to quenching.

An independent line of evidence against a dominant role for progenitor bias comes from stellar metallicity.
Quiescent galaxies are observed to have higher stellar metallicities than star-forming galaxies of the same mass, for both central and satellite populations \citep{pengMassEnvironmentDrivers2010, lyuHalosGalaxiesVII2023}, which \citet{fraser-mckelvieSAMIGalaxySurvey2022} attribute to the size difference between the two populations \citep{vanderwel3DHSTCANDELSEvolution2014, vanderwelStellarHalfmassRadii2024}, together with the anti-correlation between galaxy size and stellar metallicity \citep{sanchez-menguianoStellarMassNot2024, maRevisitingFundamentalMetallicity2024, jiaPotentialdrivenMetalCycling2025, liCentralVelocityDispersion2025, wangOriginGalaxySizestellar2026}.
Under progenitor bias, however, quiescent galaxies would inherit the metallicities characteristic of the high-redshift epoch at which they quenched, when the global galaxy population was systematically more metal-poor \citep{maiolinoAMAZEEvolutionMassmetallicity2008, kashinoStellarMassStellar2022}.
Progenitor bias would therefore predict lower stellar metallicities in quiescent galaxies relative to their present-day star-forming counterparts, in direct contradiction with observations.
Taken together, both the simulation results and this observational argument rule out progenitor bias as the dominant driver of the structural differences between star-forming and quiescent galaxies.

\subsection{Mass-size relation of star-forming and quiescent galaxies}
\label{sub:mass_size_relation_for_star_forming_and_quiescent_galaxies} 

\begin{figure}
  \begin{center}
    \includegraphics[width=0.95\linewidth]{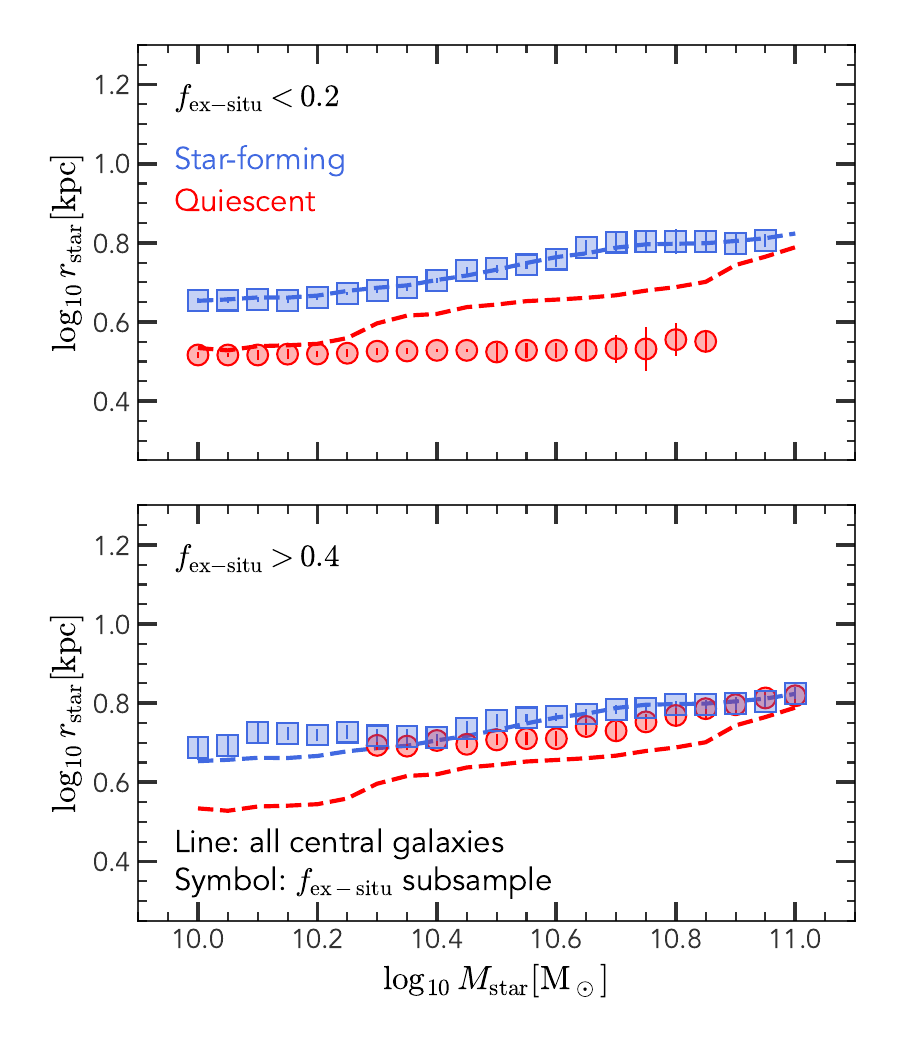}
  \end{center}
  \caption{
    Mass--size relation for star-forming (blue squares) and quiescent (red circles) central galaxies at $z=0$, split by ex-situ stellar mass fraction: merger-poor galaxies with $\fex < 0.2$ (upper panel) and merger-rich galaxies with $\fex > 0.4$ (lower panel).
    Blue and red dashed lines show the median mass--size relation for star-forming and quiescent galaxies, respectively.
    In the merger-poor sample ($\fex < 0.2$), quiescent galaxies are systematically more compact than their star-forming counterparts across the full stellar mass range, with the size deficit persisting at all masses.
    In the merger-rich sample ($\fex > 0.4$), the size difference between star-forming and quiescent galaxies is substantially reduced, with the two populations following similar mass--size relations particularly at high stellar masses.
    This contrast demonstrates that the size deficit of quiescent galaxies is driven primarily by the merger-poor population, in which compactness reflects the structural conditions that facilitate SMBH growth and quenching through internal disc instability, rather than being a universal consequence of the quenching process itself.
  }
  \label{fig:msr_fex}
\end{figure}

The size deficit of quiescent galaxies relative to star-forming counterparts at fixed stellar mass is one of the most robustly established correlations in galaxy evolution, confirmed observationally across a wide range of redshifts \citep{shenSizeDistributionGalaxies2003, vanderwel3DHSTCANDELSEvolution2014, vandokkumFormingCompactMassive2015, straatmanSizesMassiveQuiescent2015, vanderwelStellarHalfmassRadii2024, jiaSizeGrowthShort2024, songTransitionOutsideinInsideout2026} and reproduced in cosmological simulations \citep{furlongSizeEvolutionNormal2017, genelSizeEvolutionStarforming2018, ludlowEvolutionSizesAngular2026}.
The half-stellar-mass radius of quiescent galaxies is about half of that of star-forming galaxies around $M_{\rm star}\sim 10^{10}\,\msun$, and this difference persists from $z\approx 0.5$ to $z\approx 1.5$.
Understanding the physical origin of this deficit, whether it reflects conditions at the time of quenching, the quenching process itself, or subsequent evolution, remains an open question.

Fig.~\ref{fig:msr_fex} addresses this question directly by separating central galaxies into those with $f_{\rm ex-situ} < 0.2$ and $f_{\rm ex-situ}> 0.4$.
In the merger-rich subsample ($\fex > 0.4$), the mass--size relations of star-forming and quiescent galaxies are strikingly similar, particularly at high stellar masses, where mergers have driven both populations to comparable sizes.
In the merger-poor subsample ($\fex < 0.2$), by contrast, quiescent galaxies are systematically and substantially more compact than their star-forming counterparts across the mass range probed here, with the size deficit persisting and growing with stellar mass.

This result is consistent with the empirical picture proposed by \citet{vandokkumFormingCompactMassive2015}, in which the observed correlation between central stellar density and quiescence suggests that the quiescent population is a biased draw from the compact tail of the star-forming size distribution.
Our analysis extends this picture by identifying the physical mechanism responsible in the merger-poor regime: compact galaxies grow their SMBHs more efficiently through morphology-modulated accretion, as demonstrated in \S\,\ref{sub:morphology_modulated_smbh_growth}, and are therefore quenched preferentially before their star-forming counterparts of the same mass.
The size deficit in the merger-poor population is therefore a structural precondition for quenching rather than its consequence, a conclusion broadly consistent with observational evidence that compact star-forming galaxies at $z \approx 2$ are the progenitors of compact quiescent galaxies\footnote{This does not support progenitor bias, which requires the progenitors of quiescent galaxies to have the same structure as galaxies that remain star-forming. The observation is the opposite: compact star-forming galaxies are preferentially quenched, so the two populations differ before quenching begins.} \citep{barroCANDELSProgenitorsCompact2013, vandokkumFormingCompactMassive2015}.

The contrasting behaviour of the merger-rich population also carries an important implication: it shows that the size deficit is not a universal property of the quenching process, but depends fundamentally on the channel through which quenching occurs.
When mergers dominate, they increase galaxy size at fixed stellar mass regardless of star formation status, so star-forming and quiescent galaxies grow in size together and the size difference between the two populations is erased.
When internal secular processes dominate, compactness is the structural condition that enables quenching, causing the size offset between these two populations.
The size difference between star-forming and quiescent galaxies therefore encodes information about the relative importance of these two channels across the galaxy population.

\subsection{The morphology of rejuvenated galaxies}
\label{sub:reju} 

\begin{figure*}
  \begin{center}
    \includegraphics[width=0.95\linewidth]{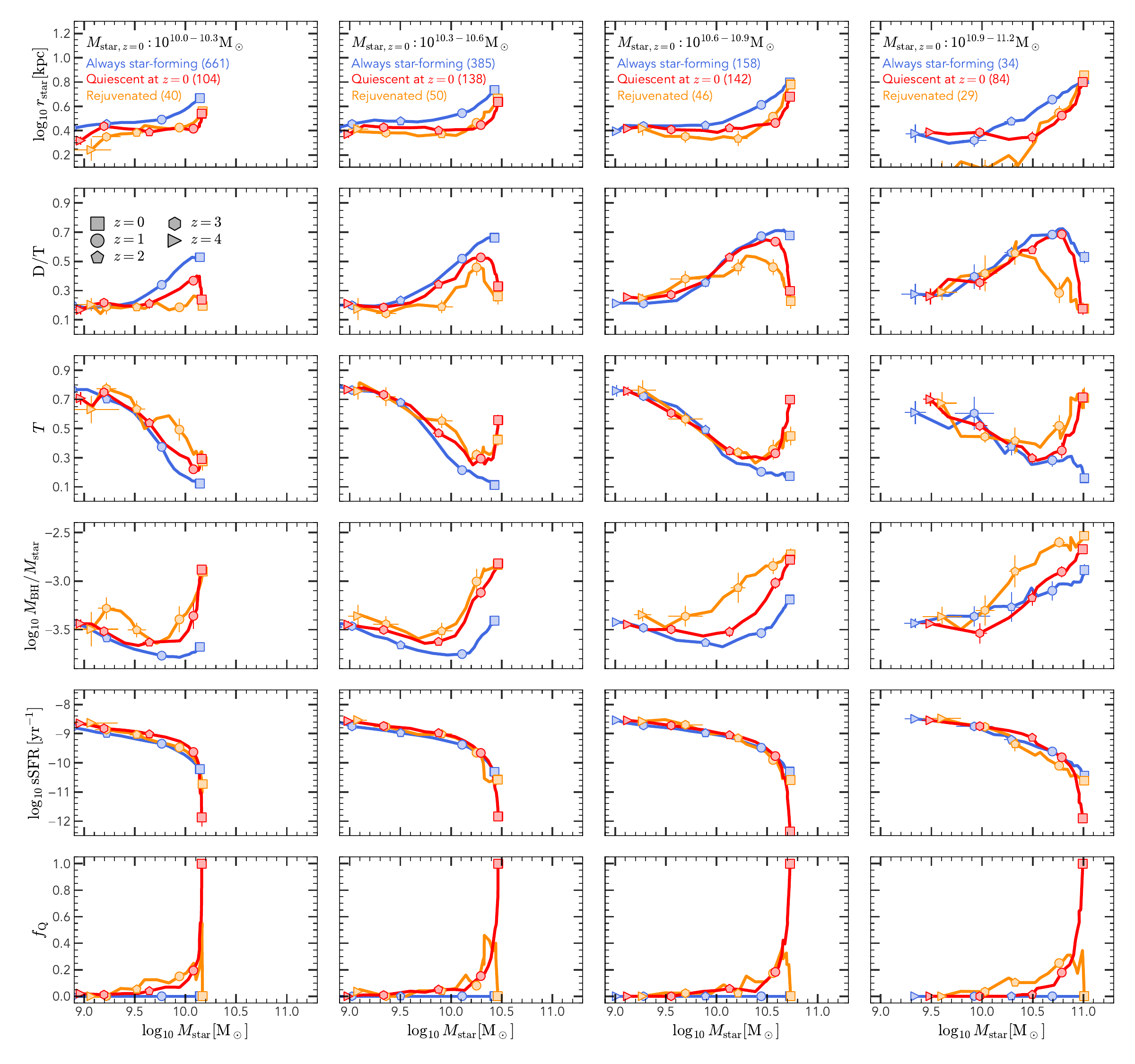}
  \end{center}
  \caption{
    As Fig.~\ref{fig:evolution_afo_mstar2}, but with the star-forming sample split into galaxies that remain star-forming throughout and galaxies that were quenched at an earlier epoch and have since resumed star formation.
  }
  \label{fig:evolution_afo_mstar_reju.pdf}
\end{figure*}

\begin{figure}
  \begin{center}
    \includegraphics[width=0.9\linewidth]{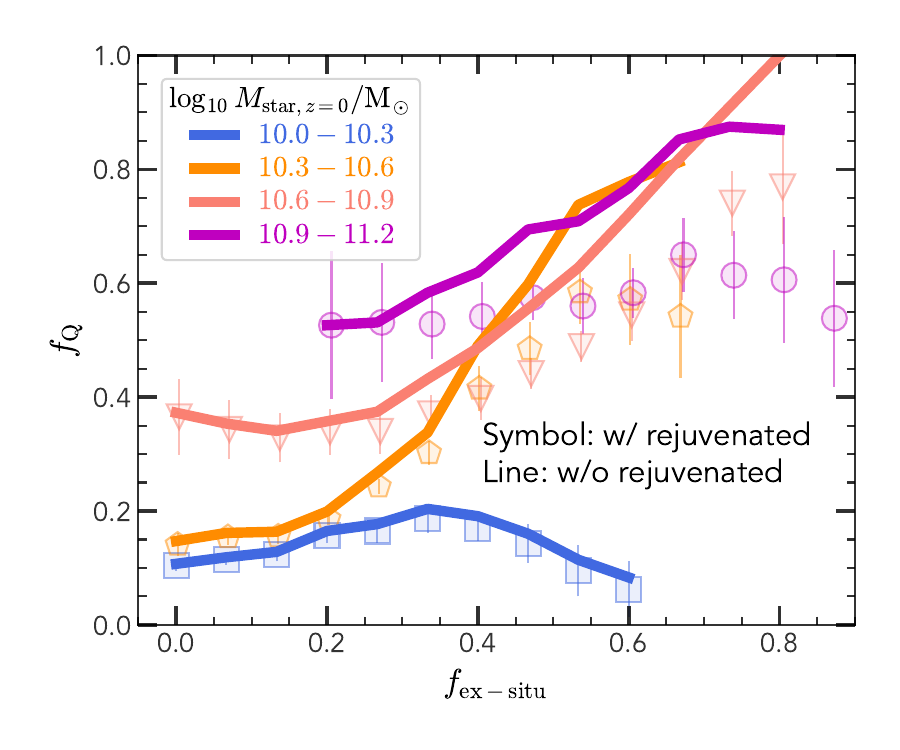}
  \end{center}
  \caption{
    As the bottom right panel of Fig.~\ref{fig:morphology_merger}, but with solid lines showing the result when rejuvenated galaxies are excluded from the star-forming sample.
    Symbols have the same meaning as in Fig.~\ref{fig:morphology_merger}.
  }
  \label{fig:morphology_merger_reju}
\end{figure}

A galaxy that is star-forming at $z=0$ may have been quenched at an earlier epoch and later resumed star formation \citep[see also][]{trayfordItNotEasy2016}.
We identify these rejuvenated galaxies by requiring that a galaxy satisfies the $z=0$ star-forming criterion and falls below the quiescence threshold in at least one earlier snapshot along its main progenitor branch.
They account for about 5 per cent of the nominally star-forming centrals in the lowest mass bin considered here, rising to about 50 per cent in the highest, so at the highest masses the star-forming sample is half composed of galaxies that have quenched before.

Fig.~\ref{fig:evolution_afo_mstar_reju.pdf} shows the evolution of the rejuvenated population alongside the star-forming and quiescent populations.
Size, disc-to-total ratio and triaxiality all separate the rejuvenated galaxies from the star-forming population.
Their sizes track the quiescent population across the full mass range, remaining below the star-forming tracks by $\approx0.1$--$0.2$ dex once the progenitors exceed $10^{9.5}\,\msun$.
Their disc-to-total ratios follow the quiescent tracks and end at $z=0$ well below the star-forming values, and their triaxialities are correspondingly higher, placing them with the quiescent population rather than with the oblate star-forming discs.
The offsets are established early and persist to $z=0$, so the resumption of star formation does not restore a star-forming structure.
The same grouping appears in $M_{\rm BH}/M_{\rm star}$, where the rejuvenated galaxies lie above the star-forming tracks at all masses, consistent with a population that has already grown the SMBH mass required for AGN feedback to be effective.

Fig.~\ref{fig:morphology_merger_reju} shows how this population affects the relation between the quiescent fraction and $\fex$.
Symbols repeat the measurement of Fig.~\ref{fig:morphology_merger}, in which rejuvenated galaxies are counted as star-forming, and solid lines show the same quantity with these galaxies removed from the star-forming sample so that $f_\mq$ is measured against galaxies that have never quenched.

The two measurements agree at low $\fex$ in all four mass bins.
The flatness of $f_\mq$ below $\fex\approx0.3$ and for the lowest-mass bin are therefore not produced by rejuvenation, and the conclusion of Section~\ref{sub:merger_driver}, that merger activity alone does not determine whether a galaxy quenches, is unaffected by how these galaxies are classified.
The two measurements separate at high $\fex$, and the separation grows with stellar mass.
Above $10^{10.3}\,\msun$ the lines rise more steeply than the symbols once $\fex\gtrsim0.4$, and in the two highest mass bins the difference at the highest $\fex$ reaches $\Delta f_\mq\approx0.2$--$0.3$.
Excluding rejuvenated galaxies therefore strengthens the dependence of $f_\mq$ on $\fex$.

\subsection{Implications for semi-analytic models of galaxy formation}
\label{sub:implications_sam}

\begin{figure}
  \begin{center}
    \includegraphics[width=0.95\linewidth]{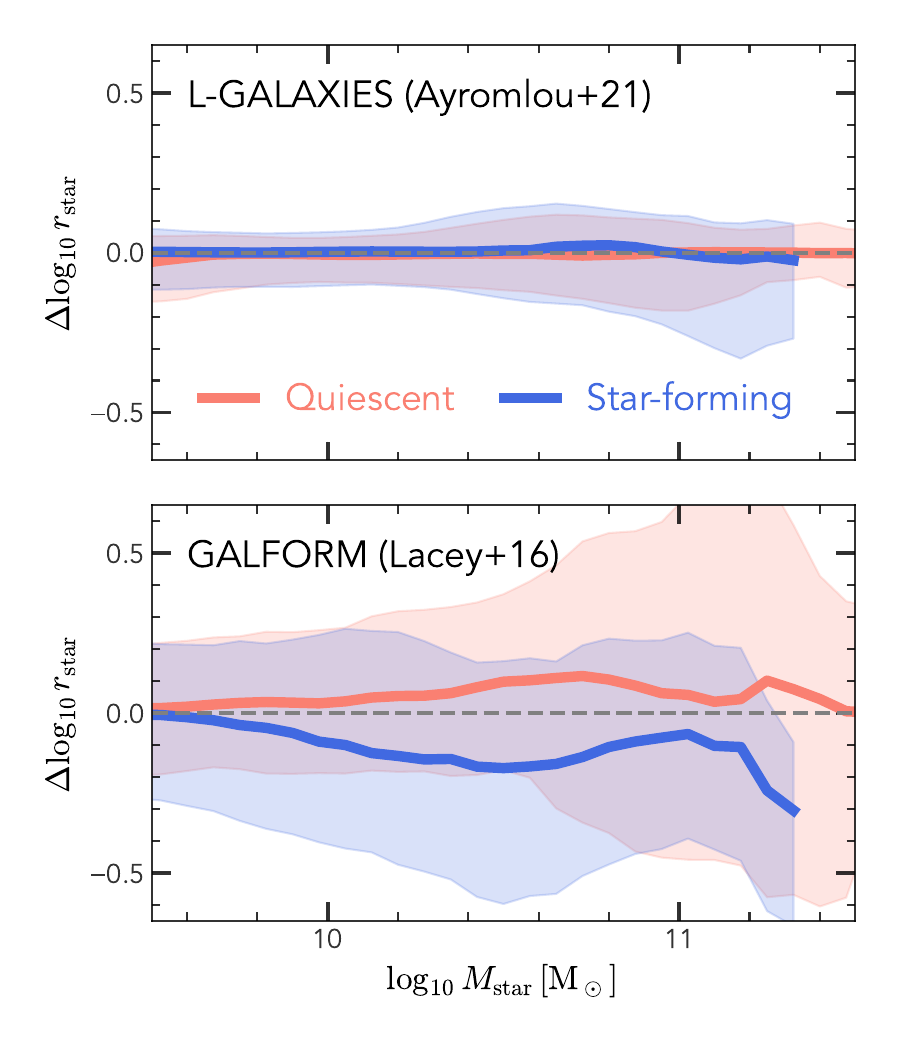}
  \end{center}
  \caption{
    Offset in stellar half-mass radius from the median mass--size relation of the full central galaxy population as a function of stellar mass, for L-GALAXIES \citep[top,][]{ayromlouGalaxyFormationLGALAXIES2021} and GALFORM \citep[bottom,][]{laceyUnifiedMultiwavelengthModel2016}.
    Lines show the median offset for quiescent and star-forming centrals, and shaded regions the 16th to 84th percentile range.
    L-GALAXIES predicts coincident populations at all masses, whereas GALFORM places quiescent galaxies above star-forming galaxies by up to $\approx0.2$ dex, the opposite trend to observational results.
  }
  \label{fig:size_diff_sam}
\end{figure}

The results presented in this paper carry implications for semi-analytic models of galaxy formation, which remain the dominant framework for making statistical predictions about galaxy populations across cosmic time \citep{coleHierarchicalGalaxyFormation2000, guoDwarfSpheroidalsCD2011, porterUnderstandingStructuralScaling2014, lagosSharkIntroducingOpen2018}.
Many of these models rely on galaxy mergers as the primary driver of both morphological transformation and SMBH growth.
L-GALAXIES grows SMBHs predominantly through this channel \citep{guoDwarfSpheroidalsCD2011, henriquesGalaxyFormationPlanck2015, ayromlouGalaxyFormationLGALAXIES2021}: a major merger destroys the disc, transfers gas to the bulge, and triggers a starburst that feeds the SMBH, which in turn drives the AGN feedback that quenches the central galaxy.
The size difference between star-forming and quiescent galaxies provides a direct test of this prescription, since a purely merger-driven route to quiescence should leave quiescent galaxies more extended than star-forming galaxies at fixed stellar mass.

Some models supplement the merger channel with secular processes as additional pathways for bulge formation and SMBH feeding.
In GALFORM \citep{coleHierarchicalGalaxyFormation2000, laceyUnifiedMultiwavelengthModel2016}, SHARK \citep{lagosSharkIntroducingOpen2018, lagosDiverseNatureFormation2022}, and the Santa Cruz model \citep{porterUnderstandingStructuralScaling2014}, galaxy disc transfers mass to the spheroid component and triggers a starburst that feeds the SMBH once its self-gravity exceeds a critical threshold under the violent disc instability \citep[e.g.][]{efstathiouStabilityMassesDisc1982}.
GAEA \citep{deluciaElementalAbundancesMilky2014, hirschmannGalaxyAssemblyStellar2016, xieH2basedStarFormation2017} models cold gas accretion onto the SMBH explicitly, following the angular momentum loss required for gas to reach the centre through both the merger and the disc instability channels \citep{fontanotRiseActiveGalactic2020}.

Fig.~\ref{fig:size_diff_sam} shows the size offset of star-forming and quiescent central galaxies from the median mass--size relation of the full central population, for the L-GALAXIES implementation of \citet{ayromlouGalaxyFormationLGALAXIES2021} and the GALFORM implementation of \citet{laceyUnifiedMultiwavelengthModel2016}.
The top panel shows that L-GALAXIES predicts no size difference between the two populations at any stellar mass \citep[see also][]{vaniProbingGalaxyEvolution2025}, and so fails to reproduce the observed result that quiescent galaxies are more compact than star-forming galaxies \citep[e.g.][]{vanderwelStellarHalfmassRadii2024}.
Identifying why quiescent galaxies in this model are not larger than star-forming galaxies, as a merger-driven quenching scenario would predict, is beyond the scope of this paper.
The lower panel shows that the violent disc instability channel in GALFORM does not resolve the discrepancy: quiescent galaxies are more extended than star-forming galaxies across most of the mass range.

The observed size difference between star-forming and quiescent galaxies is a stringent test of how galaxy formation models treat the processes that drive central galaxies to quiescence.
Reproducing this difference is not a trivial task for either semi-analytic models or hydrodynamical simulations \citep[SIMBA provides a counter-example among hydrodynamical simulations,][]{daveSIMBACosmologicalSimulations2019}, particularly where SMBH growth is not resolved and instead follows a subgrid prescription.
The difficulty is itself informative, because every model of this kind follows the structural evolution of galaxies alongside the evolution of the scaling relations, and so contains progenitor bias by construction (see \S~\ref{sub:Progenitor bias} for details).
If the inheritance of structural properties at the epoch of quenching were sufficient to produce the present-day offset, every model would reproduce it, and only a few do.
The offset therefore requires a physical connection between galaxy structure and quenching that the models must implement explicitly, and our results point to a route for doing so by relating SMBH growth to galaxy morphology, so that compact galaxies build their black holes faster and quench earlier at fixed stellar mass.
Whether such a prescription reproduces the observed ordering while preserving the scaling relations these models are already calibrated against remains to be tested.

\section{Summary}
\label{sec:summary}

The tight correlation between galaxy morphology and star formation activity is one of the most fundamental observational facts of galaxy evolution, yet the physical processes responsible for it remain widely debated.
In this paper we have used the EAGLE cosmological hydrodynamical galaxy formation simulation to study the co-evolution of galaxy morphology and star formation quenching, tracing central galaxies from $z \gtrsim 4$ to $z = 0$.
By characterising morphology through size, disc-to-total ratio, and triaxiality, and by using the ex-situ stellar mass fraction as a proxy for merger history, we have decomposed the morphology--quenching connection into physically distinct channels and identified the structural conditions that govern galaxy quenching.
Our main findings are as follows.

\begin{enumerate}

  \item
        The progenitors of $z=0$ star-forming and quiescent galaxies follow indistinguishable disc-to-total ratio ($\rm D/T$) and triaxiality ($T$) tracks at high redshift, diverging only below $z\approx 1$ alongside the onset of quenching, whereas the size and $M_{\rm BH}/M_{\rm star}$ divergences are established earlier, from $z\approx2$.
        By $z=0$ the quiescent population reaches $T\approx0.3$--$0.7$ and ${\rm D/T}\approx0.1$--$0.3$ against $T\approx0.1$--$0.2$ and ${\rm D/T}\approx0.5$--$0.7$ for star-forming galaxies, with smaller sizes and more massive SMBHs at fixed stellar mass (Figs.~\ref{fig:evolution_afo_mstar1} and \ref{fig:evolution_afo_mstar2}).

  \item
        We quantify the cumulative merger history of each galaxy by $\fex$, the fraction of its stellar mass that formed outside the main progenitor branch, and find that mergers drive a coherent co-evolution of morphology, SMBH growth and quenching.
        At fixed stellar mass, galaxies with higher $\fex$ have larger $r_{\rm star}$, lower $\rm D/T$, higher $T$, and host more massive SMBHs.
        This establishes the role of mergers in linking morphological transformation to quenching by promoting the SMBH growth that drives AGN-feedback quenching.
        For galaxies with $M_{\rm star} \leq 10^{10.3}\,\msun$, the quiescent fraction is insensitive to $\fex$, showing that mergers affect quenching only indirectly through SMBH growth rather than quenching galaxies directly (Fig.~\ref{fig:morphology_merger}).

  \item
        Controlling for both stellar mass and $\fex$, we still find structural differences between star-forming and quiescent galaxies, so a mechanism beyond mergers must operate (Fig.~\ref{fig:morphology_quenching_fexsitu}).
        At fixed stellar mass and $\fex$, galaxy morphology modulates SMBH growth throughout the star-forming phase: $\rm D/T$ correlates negatively with $M_{\rm BH}$ in every $\fex$ bin, and the size--$M_{\rm BH}$ correlation is negative at low $\fex$ and reverses sign at high $\fex$, so compactness drives preferential SMBH growth and quenching where mergers are absent (Fig.~\ref{fig:correlation_to_mbh}).

  \item
        Compactness also governs quenching at fixed stellar and SMBH mass: galaxies destined to quench are more compact by $\approx0.1$--$0.2$ dex than star-forming galaxies of the same stellar and SMBH mass, while their disc-to-total ratios are indistinguishable, identifying compactness rather than the loss of rotational support as the structural condition that facilitates AGN-feedback-driven quenching (Fig.~\ref{fig:quenching_morphology_role}).

  \item
        Morphological transformation does not cease at quenching but continues through internal secular processes.
        Low-$\fex$ quiescent galaxies grow in size by $\approx$0.2 dex, decrease their disc-to-total ratio by $\approx$0.2--0.4, and increase their triaxiality by $\approx$0.1--0.4 between $z_\mq$ and $z = 0$, at negligible stellar mass growth, with the magnitude of these changes increasing with $z_\mq$ (Fig.~\ref{fig:morphology_evolution_after_quenching}).

  \item
        Compact, disc-dominated low-$\fex$ quiescent galaxies undergo a near-vertical transformation on the $M_{\rm star}-{\rm D/T}$ and $M_{\rm star}-T$ planes at approximately constant stellar mass, driven by internal disc instability that converts oblate discs into prolate bars with significantly enlarged galaxy size and suppressed rotational support (Figs.~\ref{fig:evolution_fex_subsample} and \ref{fig:example1}).
        This transformation produces a characteristic horizontal shift in the $b/a$--$c/a$ shape plane that is geometrically distinct from the triaxial configurations produced by mergers (Fig.~\ref{fig:prolate_distribution.pdf}), and is consistent with the \citet{efstathiouStabilityMassesDisc1982} disc instability criterion (Fig.~\ref{fig:fstar}).

  \item
        The $z=0$ prolate population is supplied by two channels of comparable importance at intermediate mass.
        Near $M_{\rm star}\approx10^{10.6}\,\msun$, close to half of all prolate galaxies have $\fex<0.2$ and are therefore products of secular disc instability rather than mergers, with mergers dominating at higher masses (Fig.~\ref{fig:prolate_demographic}).

  \item
        The structural differences between star-forming and quiescent galaxies are in place before quenching and comparable in magnitude to their $z=0$ values, ruling out progenitor bias as their dominant origin.
        The residual difference between the two epochs is accounted for by the post-quenching secular evolution above (Fig.~\ref{fig:progenitor_bias.pdf}).

  \item
        The size deficit of quiescent galaxies is carried by the low-$\fex$ population alone: high-$\fex$ galaxies follow the same mass--size relation whether or not they quench, so the size offset is a signature of the merger-poor quenching channel rather than of quenching in general (Fig.~\ref{fig:msr_fex}).

  \item
        Rejuvenated galaxies, which are star-forming at $z=0$ but were quenched at an earlier epoch, are structurally indistinguishable from the quiescent population in size, $\rm D/T$, $T$ and $M_{\rm BH}/M_{\rm star}$, so the resumption of star formation does not restore a star-forming structure (Fig.~\ref{fig:evolution_afo_mstar_reju.pdf}).
        Removing them from the star-forming sample strengthens the dependence of $f_\mq$ on $\fex$ (Fig.~\ref{fig:morphology_merger_reju}).

  \item
        Semi-analytic models that grow SMBHs through mergers and disc instabilities alone predict either no size difference between star-forming and quiescent centrals or one of the wrong sign (Fig.~\ref{fig:size_diff_sam}).
        Reproducing the observed difference requires SMBH accretion to depend on the structural state of the galaxy, making this measurement a stringent test of how galaxy formation models connect structure to quenching.

\end{enumerate}

Together, these results reveal that the morphology--quenching connection in EAGLE is not a monolithic consequence of a single physical process, but the superposition of at least four distinct channels operating across different mass and merger regimes.
More importantly, they rule out progenitor bias as the driver of this connection, and establish instead a physical link between structural evolution and quenching.
These findings also carry important implications for the semi-analytic modelling of galaxy formation, calling for a more careful treatment of SMBH growth and the quenching of central galaxies.

\section*{Acknowledgements}

KW thanks the inspiring discussions with Rob Crain, Andrew Pontzen, Joop Schaye, Matthew Frosst, Hiranya Peiris, Jie Wang, Yong Shi, Shude Mao, Haonan Zheng, and Yangyao Chen at different stages of this work.
KW acknowledges the use of Claude (Anthropic) as a writing and research aid in the preparation of this manuscript, including literature searches, drafting assistance, and language editing; all scientific content, analysis, and conclusions are the authors' own.

This work is supported by the Science and Technologies Facilities Council (STFC) through grant ST/X001075/1.
SB is supported by the UK Research and Innovation (UKRI) Future Leaders Fellowship [grant number MR/V023381/1 and UKRI2044].
This work is co-funded by the European Union (Widening Participation, ExGal-Twin, GA 101158446). Views and opinions expressed are however those of the author(s) only and do not necessarily reflect those of the European Union. Neither the European Union nor the granting authority can be held responsible for them.
Y.P. acknowledges support from the National Natural Science Foundation of China (NSFC) under grant Nos. 12125301 and 12192222, and from the New Cornerstone Science Foundation through the XPLORER PRIZE.

This work used the DiRAC@Durham facility managed by the Institute for Computational Cosmology on behalf of the STFC DiRAC HPC Facility
(www.dirac.ac.uk).
The equipment was funded by BEIS capital funding via STFC capital grants ST/K00042X/1, ST/P002293/1, ST/R002371/1 and ST/S002502/1, Durham University and STFC operations grant ST/R000832/1.
DiRAC is part of the National e-Infrastructure.

\section*{Data availability}

The data underlying this article will be shared on reasonable request to the corresponding author.

\bibliographystyle{mnras}
\bibliography{bibtex.bib}

\appendix

\section{The importance of galaxy mergers}
\label{sec:the_impact_of_galaxy_mergers}

\begin{figure*}
  \begin{center}
    \includegraphics[width=0.95\linewidth]{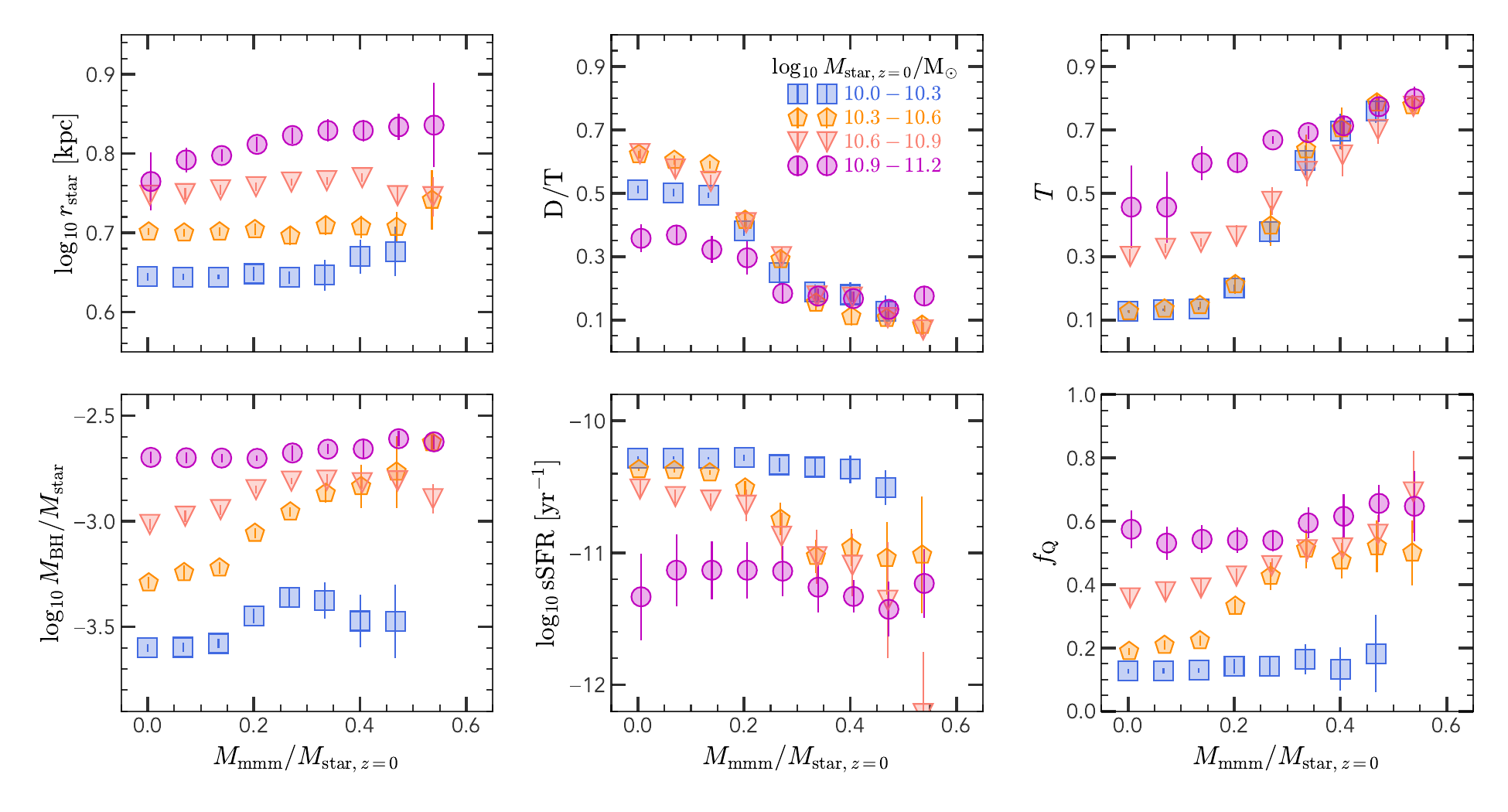}
  \end{center}
  \caption{
    Same as Fig.~\ref{fig:morphology_merger}, except that the $x$-axis shows the stellar mass ratio between the most massive merger along the main branch and the $z=0$ descendant galaxy, $M_{\rm mmm}/M_{{\rm star},z=0}$, rather than the ex-situ stellar mass fraction, $\fex$.
    All trends qualitatively agree with Fig.~\ref{fig:morphology_merger}, but the overall change in galaxy properties with $M_{\rm mmm}/M_{{\rm star},z=0}$ is small, reflecting that the contribution from multiple mergers, rather than a single merger event, impacts galaxy properties and evolution more.
  }
  \label{fig:morphology_merger2}
\end{figure*}

Fig.~\ref{fig:morphology_merger2} tests whether the trends identified in \S\,\ref{sub:merger_driver} reflect the cumulative effect of merger activity or are dominated by a single major merger event.
We repeat the analysis of Fig.~\ref{fig:morphology_merger}, replacing $\fex$ with $M_{\rm mmm}/M_{{\rm star},z=0}$, the stellar mass contributed by the single most massive merger along each galaxy's main branch, normalised by its $z=0$ stellar mass.
All six properties, $r_{\rm star}$, D/T, $T$, $M_{\rm BH}/M_{\rm star}$, sSFR, and $F_{\rm Q}$, vary with $M_{\rm mmm}/M_{{\rm star},z=0}$ in the same sense as with $\fex$: galaxies whose most massive merger contributed a larger mass fraction are systematically larger, less disc-dominated, more triaxial, host more massive SMBHs, and are more likely to be quiescent.
The dependence on $M_{\rm mmm}/M_{{\rm star},z=0}$, however, is weaker than the corresponding dependence on $\fex$ in Fig.~\ref{fig:morphology_merger}.
This weaker dependence confirms that the morphological and structural transformation established in \S\,\ref{sec:the_impact_of_galaxy_mergers} is not driven by a single dominant merger, but accumulates over a galaxy's full merger history, consistent with our use of $\fex$ as the primary merger-activity proxy throughout the main text.
The galaxy size trend shows the clearest example of this: the dependence of $r_{\rm star}$ on $\fex$ is much stronger than its dependence on $M_{\rm mmm}/M_{{\rm star},z=0}$, indicating that a single merger event changes galaxy size by only a small amount, consistent with \citet{wang26}, who found that a single merger event increases galaxy size by $\approx 0.03-0.1$ dex.

\bsp  
\label{lastpage}
\end{document}